# Meta-Analysis of Gene Level Association Tests

Dajiang J. Liu[1*], Gina M. Peloso[2,3*], Xiaowei Zhan[1*], Oddgeir Holmen[4,*] Matthew Zawistowski[1], Shuang Feng[1], Majid Nikpay[5], Paul L. Auer[5], Anuj Goel[6,10], He Zhang[7,8], Ulrike Peters[5,9], Martin Farrall[6,10], Marju Orho-Melander[11], Charles Kooperberg[5,12], Ruth McPherson[13], Hugh Watkins[6,10], Cristen J. Willer[7,8], Kristian Hveem[4], Olle Melander[11], Sekar Kathiresan[2,14+] and Gonçalo R. Abecasis[1,+]

[1]Center for Statistical Genetics, Department of Biostatistics, University of Michigan School of Public Health, Ann Arbor, MI 48109; [2]Broad Institute of Harvard and MIT, Cambridge, MA; [3]Center for Human Genetic Research and Cardiovascular Research Center, Massachusetts General Hospital, Boston, MA; [4]Department of Public Health and General Practice, Norwegian University of Science and Technology, Trondheim 7489, Norway; [5]Public Health Sciences Division, Fred Hutchinson Cancer Research Center, Seattle WA 98109, USA; [6]Wellcome Trust Centre for Human Genetics, University of Oxford, Oxford OX3 7BN, United Kingdom; [7]Division of Cardiology, Department of Internal Medicine, University of Michigan Medical School, Ann Arbor, MI 48109; [8]Department of Human Genetics, University of Michigan Medical School, Ann Arbor, MI 48109; [9]Department of Epidemiology, University of Washington School of Public Health, Seattle, WA; [10]Department of Cardiovascular Medicine, University of Oxford, Oxford, UK; [11]Department of Clinical Sciences, Lund University, Malmö, Sweden; [12]Department of Biostatistics, University of Washington School of Public Health, Seattle, WA; [13]University of Ottawa Heart Institute, Ottawa, Ontario, Canada; [14]Harvard Medical School, Cambridge, MA

* These authors contributed equally and should be considered joint first authors.

+ These authors jointly directed the study.

**Correspondence To:**

Dajiang J. Liu (dajiang@umich.edu)

Or

Gonçalo R. Abecasis (goncalo@umich.edu)

Center for Statistical Genetics,

Department of Biostatistics,

1415 Washington Heights,

Ann Arbor, MI 48109

**Abstract**

**The vast majority of connections between complex disease and common genetic variants were identified through meta-analysis, a powerful approach that enables large samples sizes while protecting against common artifacts due to population structure, repeated small sample analyses, and/or limitations with sharing individual level data. As the focus of genetic association studies shifts to rare variants, genes and other functional units are becoming the unit of analysis. Here, we propose and evaluate new approaches for meta-analysis of rare variant association. We show that our approach retains useful features of single variant meta-analytic approaches and demonstrate its utility in a study of blood lipid levels in ~18,500 individuals genotyped with exome arrays.**

Proceeding from the discovery of a genetic association signal to a mechanistic insight about human biology should be much easier for alleles with clear functional consequence, including non-synonymous, splice altering and protein truncating alleles. Most of these alleles are very rare, with only one such allele expected to reach MAF>5% in the average human gene[1]. Recent advances in exome sequencing and the development of exome genotyping arrays are enabling explorations of the very large reservoir of rare coding variants in humans and are expected to accelerate the pace of discovery in human genetics[2].

Rare variants can be examined using association tests that group alleles in a gene or other functional unit[3]. Compared to tests of individual alleles, this grouping can increase power, especially when applied to large samples where several rare variants are observed in the same functional unit[4]. The simplest rare variant tests consider the number of potentially functional alleles in each individual[5], but the tests can be refined to weigh variants according to their likely functional impact[6], to allow for imputed or uncertain genotypes[7,8], or to allow variants that increase and decrease risk to reside in the same gene[9-11] (a feature that is important when the same gene harbors hypermorph and hypomorph alleles[12]). The optimal strategy for grouping and weighting rare variants – ranging from focusing on protein truncation alleles to examining all non-synonymous variants and encompassing strategies that examine all variants with frequency <5% as well as alternatives that examine only singletons – depends on the unknown genetic architecture of each trait and each locus.

Here, we describe practical approaches for meta-analysis of rare variants. Our approach starts with simple statistics that can be calculated in an individual study. We then show that, when these statistics are shared, a wide variety of gene-level association tests can be executed centrally – including both weighted or un-weighted burden tests with fixed[5] or variable frequency threshold[6] and sequence kernel association tests (SKAT) that accommodate alleles with opposite effects within a gene[9]. Our approach generates comparable results to sharing individual level data (and, in fact, identical results when allowing for between study heterogeneity in nuisance parameters, such as trait means, variances and covariate effects). As an illustration of our approach, we analyze blood lipid levels in >18,500 individuals genotyped with exome genotyping arrays. Our analysis of blood lipid levels provides examples of loci where signal for gene-level association tests exceeds signal for single variant tests and shows that our approach can recover signals driven by very rare variants (frequency <0.05%). Given that very large sample sizes are required for successful rare variant association studies, we expect our methods (and refined versions thereof) will be widely useful.

Our approach is based on the insight that analogues of most gene level association tests can be constructed using single variant test statistics and knowledge of linkage disequilibrium relationships. As shown in **Online Methods,** simple[13] and weighted[10,14] burden tests, variable threshold tests[6] and tests allowing for variants with opposite effects[9] can be constructed in this manner. We meta-analyze single variant statistics using the Cochran-Mantel-Haenszel method, calculate variance-covariance matrices for these statistics, and construct gene-level association tests by combining the two. In **Supplementary Material**, we show that rare variant statistics generated in this way identical to those obtained by sharing individual level data and allowing for heterogeneity in nuisance parameters, with no loss of power. As in other meta-analysis settings, sharing summary statistics accelerates the overall analysis process, mitigates concerns about participant confidentiality, and reduces the risk that data will be used for unapproved analyses (as always, to avoid violating the trust of research subjects, we strongly recommend that investigators sharing summary statistics agree that these will not be used to identify research subjects). For evaluating significance, we propose methods for calculating p-values using asymptotics and also Monte-Carlo methods that use knowledge of linkage disequilibrium relationships to sample plausible combinations of single variant statistics and then generate empirical distributions for our gene-level statistics. Since asymptotic p-values rely on numerical integration to evaluate high-dimensional integrals that can be numerically unstable, Monte-Carlo methods can be used to verify interesting p-values.

We first evaluated our method using simulations. Genes were simulated as stretches of 5,000 base-pairs using the coalescent[15] and a demographic model (including an ancient bottleneck, recent exponential growth, differentiation and migration) calibrated to mimic a sample of multiple European populations[16,17] (**Supplementary Figure 1** and **Supplementary Material**). The simulations produced samples of 1,000 individuals, each from several related populations, typically including a few shared variants and many population specific variants – as expected when the distribution of rare variants is geographically restricted[18]. Half of the simulated variants were randomly set to increase trait values by $1/8^{th}$ of a standard deviation (see **Supplementary Figure 2 and 3** for similar results using alternative trait models).

We analyzed each simulated sample with a series of gene-level association tests. **Figure 1** compares results obtained for 10,000 simulated genes using our meta-analysis approach to a combined analysis of individual level data across studies. For variable threshold tests, we found the p-values were sometimes slightly different ($r^2$=0.995 between the two sets of log p-values); for the other two tests p-values and test statistics were indistinguishable. Calculation of analytical p-values for variable threshold tests requires the evaluation of high-dimensional integrals that can be numerically unstable and is thus very sensitive to small differences in the variance-covariance matrix. In practice, it will often be a good idea to confirm significant p-values using our Monte-Carlo approach.

To evaluate our Monte-Carlo approach, we compared its empirical p-values to those obtained by permuting phenotypes between individuals within each study. We implemented adaptive versions of both algorithms[19], with more simulations carried out when the p-value is small and fewer simulations when the p-value is large. Log p-values for the two approaches are highly concordant ($r^2$=0.996).

We next evaluated the type I error and power under a variety of models. **Supplemental Table 1** shows that type I error is well controlled. **Supplemental Figures 4 and 5** summarize power for various simulation scenarios and meta-analysis of 3 to 100 samples of 1000 individuals each. Power always increases as more studies are analyzed. It is clear that, for the effect sizes simulated here, very large samples may be required. In some settings, power only reaches ~60% in analyses of ~100,000 individuals. In addition, we did not find a universally most powerful method, emphasizing the value of implementing a diverse set of test statistics. **Supplemental Figure 6** shows that, as expected, our meta-analysis of summary statistics results in the same power as joint-analysis of individual data. **Supplemental Figure 7** shows our approach outperforms simpler meta-analysis methods, such as those based on Fisher's method for combining p-values, in an analysis of three equal sized samples; the advantages of our method increase when studies are unequal in size and/or the number of studies rises.

We proceeded to a meta-analysis of blood lipid levels in 18,699 individuals of European ancestry genotyped with Illumina Exome arrays and drawn from 7 studies: the Women's Health Initiative[20], the Ottawa Heart Study[21], the Malmö Diet and Cancer Study – Cardiovascular Cohort (MDC)[22], the PROCARDIS Precocious Coronary Artery Disease Case Series, PROCARDIS Control series[23] and the Nord-Trøndelag Health Study (HUNT) myocardial infraction cases and matched controls[24] (see **Supplemental Table 2 and 3** for summary statistics for each of these samples, including basic demographics, summaries of lipid levels, number of non-synonymous and loss-of-function variants per individual and of variants sites shared across different studies). Overall, 171,193 variants were polymorphic in at least one individual. Among these variants, 125,702 – the vast majority – have frequency <1%.

To verify the soundness of our approach, we first compared joint analysis of individual level data to meta-analysis by splitting the largest sample (MDC) into 4 sub-samples. To mimic our meta-analysis protocol, we normal transformed trait values in each sub-sample and generated summary statistics, which were then meta-analyzed and used for gene level tests. We carried out a parallel analysis examining all individual level data jointly. Reassuringly, p-values from the two approaches are highly concordant with $r^2>0.99$ (**Supplemental Figure 8**).

We then proceeded to meta-analyze single variant association test results. The resulting test statistics appear well calibrated, with genomic control value <1.05 for all three traits (**Supplemental Figure 9**). At a significance threshold of $p<3x10^{-7}$ (corresponding to 0.05 / 171,193), we found significantly associated variants (with MAF<5%) at *LPL*, *ANGPTL4*, *LIPG*, *CD300LG*, *LIPC*, *APOB*, *HNF4A* for HDL; PCSK9, *BCAM-CBLC-PVR* (neighboring APOE), and *APOB* for LDL; *ANGPTL4, LPL, APOB,* and *MAP1A* for TG (**Supplemental Table 4**). Except for the variants in *LIPC* and *APOB,* all other significantly associated variants have frequency of >1% reflecting the limited power of single variant association tests for rare alleles.

We next carried out gene-level tests. Again, test statistics appear well calibrated, with genomic control value <1.05 (**Supplemental Figure 10**). At a significance threshold of $p<3.1x10^{-6}$ (corresponding to 0.05 / 16,153 and thus allowing for the number of genes tested), we observed association at *ANGPTL4*, LIPC, *LIPG*, *HNF4A*, *CD300LG* and *LPL* for HDL, at the *APOE*-locus (as well as nearby genes *PVR*, *BCAM*, and *CBLC*), *PCSK9* and *LDLR* for LDL, and at *ANGPTL4, APOB* and *LPL* for triglycerides (**Table 1**). **Supplemental Table 5** emphasizes that, at these loci, much stronger signals are identified in meta-analysis than in any component study. Reassuringly, these signals point to loci identified in previous genome-wide association studies and/or re-sequencing studies. Importantly, note that our approach was able to appropriately identify the signal in *LDLR* which is driven by several very rare variants (each with frequency < .00052) that nearly always increase blood LDL cholesterol levels and that, at several other loci, gene-level p-values exceeded the best single variant p-value in the gene (**Supplemental Table 6**).

An added convenience of sharing single-variant statistics together with their covariance matrices, as we propose, is that it facilitates conditional analyses, extending an idea used by Yang et al[25] for analysis of common variants in GWAS meta-analysis. **Supplemental Figure 8** illustrates how, in simulations, common variants can generate shadow rare variant association signals at nearby genes, and how our method for conditional analysis resolves the problem. In real data, we re-examined two loci of the LDL associated loci in detail, *LDLR* and *APOE-BCAM-CBLC-PVR*. For *LDLR*, we examined the relationship between rare variant signals and three nearby common variants[26]. Specifically, we conditioned on genotypes for 3 common variants (rs6511720, rs2228671 and rs72658855) exhibiting significant association in the region, and found that *LDLR* rare variant association remains significant (p-value $4.6\times10^{-7}$) (**Supplemental Table 7**). For the *APOE-BCAM-CBLC-PVR* locus, after conditioning on the common variant showing strongest association in the region (rs7412), gene-level associations at *BCAM*, *CLBC* and *PVR* become non-significant, suggesting that these rare-variant signals are the result of regional linkage equilibrium with more common and well described variants in *APOE* (**Supplemental Table 8**). For completeness, **Supplemental Figure 12 and 13** show that conditional analyses using individual level data of MDC and conditional meta-analyses of 4 sub-samples give highly concordant ($r^2>0.99$).

Our methods are implemented freely available software, including programs for calculating summary statistics, annotating the resulting summaries, performing meta-analysis and calculating gene-level statistics, and executing conditional analyses. Our tools work with standard VCF files[10,27] and Merlin format files[28].

Meta-analysis has facilitated many discoveries in common variant association studies. Here, we describe a powerful framework for meta-analysis of rare variants at the level of genes or other functional units. Through simulation and empirical evaluation, we demonstrate that our approach is well calibrated and provides comparable power to more cumbersome analyses that require pooling all individual level data. Through the analysis of blood lipids levels across seven studies, we show that our approach can detect rare variant association signals at known candidate loci. We envision that this approach will facilitate the large sample sizes required to accelerate new discoveries in complex trait genetics.

**Acknowledgements:**

The authors would like to thank Drs. Michael Boehnke, Xiaoquan (William) Wen, and Sebastian Zoellner for helpful discussions. This work was supported by research grants from the National Human Genome Research Institute, the National Eye Institute and the National Heart, Lung and Blood Institute. Gina M. Peloso was supported by Award Number T32HL007208 from the National Heart, Lung, and Blood Institute. The content is solely the responsibility of the authors and does not necessarily represent the official views of the National Heart, Lung, and Blood Institute or the National Institutes of Health. The WHI program is funded by the National Heart, Lung, and Blood Institute, National Institutes of Health, U.S. Department of Health and Human Services through contracts N01WH22110, 24152, 32100-2, 32105-6, 32108-9, 32111-13, 32115, 32118-32119, 32122, 42107-26, 42129-32, and 44221. This manuscript was prepared in collaboration with investigators of the WHI, and has been approved by the WHI. WHI investigators are listed at

http://www.whiscience.org/publications/WHI_investigators_shortlist.pdf.

For full list of PROCARDIS acknowledgements, visit www.procardis.org.

**Online Resources:**

Author's website at http://genome.sph.umich.edu/

**Figure legend**

Figure 1: Scatter plot of statistics and p-values for a simple burden test grouping variants with MAF < 1% (burden-1), variable threshold tests (VT) and tests allowing for variants with opposite effects (SKAT) in meta-analysis and joint-analysis of individual data. Three samples of 1000 European-ancestry individuals were simulated. Traits were simulated assuming that 50% of the variants in the gene region are causal and that each causal variant increases trait means by 0.125 standard deviations. Empirical p-values for the VT were obtained using our Monte-Carlo procedure to generate replicates until 100 simulated statistics exceeded the original observation or 40,000,000 statistics were simulated.

**Table 1:** Results for meta-analysis of gene-level rare variant association test. Associations that attain exome-wide significance ($p < 3.1\times10^{-6}$) are displayed. Five gene-level association tests were used to analyze the data: simple burden tests with 1% or 5% cutoff (Burden-1 and Burden-5), SKAT tests with 1% or 5% cutoff (SKAT-1 and SKAT-5) and variable threshold (VT) tests that analyze variants with MAF<5%. Significant p-values for each test are displayed in bold font. For the associations that are significant, estimates of average genetic effect are also shown.

| Gene | Gene Position[a] | Burden-1 | Burden-5 | SKAT-1 | SKAT-5 | VT | MAF Cutoff | Direction of Single Variant Association Statistics[b] | Estimates of Genetic Average Effect (s.d units) for Rare Variants under Different MAF Thresholds | | |
|---|---|---|---|---|---|---|---|---|---|---|---|
| | | | | | | | | | 0.01 | 0.05 | VT |
| | | | | | HDL | | | | | | |
| *LIPC* | chr15:58.7Mb | **$1.4\times10^{-12}$** | **$3.5\times10^{-7}$** | **$1.8\times10^{-9}$** | $1.4\times10^{-2}$ | **$4.5\times10^{-12}$** | $3.7\times10^{-3}$ | -++++--+- | 0.5 | 0.1 | 0.5 |
| *LPL* | chr8:19.8Mb | $9.7\times10^{-1}$ | **$2.5\times10^{-24}$** | $3.5\times10^{-1}$ | **$5.0\times10^{-13}$** | **$1.5\times10^{-23}$** | $2.5\times10^{-2}$ | (-)-(-)+-++ | - | -0.3 | -0.3 |
| *ANGPTL4* | chr19:8.4Mb | $2.2\times10^{-2}$ | **$2.9\times10^{-19}$** | $2.2\times10^{-2}$ | **$3.0\times10^{-19}$** | **$1.8\times10^{-18}$** | $2.6\times10^{-2}$ | (+)--++-+++ | - | 0.3 | 0.3 |
| *LIPG* | chr18:47.1Mb | $2.2\times10^{-5}$ | **$6.4\times10^{-19}$** | $2.1\times10^{-5}$ | **$2.9\times10^{-9}$** | **$4.4\times10^{-18}$** | $1.3\times10^{-2}$ | -++----(+)+ | - | 0.4 | 0.4 |
| *HNF4A* | chr20:43.0Mb | $7.5\times10^{-1}$ | **$2.8\times10^{-7}$** | $6.8\times10^{-1}$ | **$2.5\times10^{-7}$** | **$1.5\times10^{-6}$** | $4.1\times10^{-2}$ | (-)--+-+ | - | -0.1 | -0.1 |
| *CD300LG* | chr17:41.9Mb | $4.9\times10^{-1}$ | **$8.5\times10^{-7}$** | $5.2\times10^{-1}$ | $1.0\times10^{-5}$ | **$3.1\times10^{-6}$** | $3.3\times10^{-2}$ | (-)+-(+) | - | -0.1 | - |
| | | | | | LDL | | | | | | |
| *PCSK9* | chr1:55.5Mb | $1.8\times10^{-2}$ | **$7.4\times10^{-19}$** | $8.1\times10^{-2}$ | **$5.5\times10^{-17}$** | **$2.0\times10^{-28}$** | $1.3\times10^{-2}$ | (-)--(-)--+-++- | - | -0.3 | -0.5 |
| *BCAM* | chr19:45.3Mb | $1.7\times10^{-1}$ | **$1.6\times10^{-18}$** | $1.5\times10^{-1}$ | $3.0\times10^{-5}$ | **$2.6\times10^{-17}$** | $3.6\times10^{-2}$ | (-)+++(-)+-+++---+(-)+--+--++ | - | -0.1 | -0.1 |
| *CBLC* | chr19:45.3Mb | $9.4\times10^{-1}$ | **$2.0\times10^{-15}$** | $4.4\times10^{-1}$ | $1.5\times10^{-4}$ | **$1.0\times10^{-14}$** | $4.4\times10^{-2}$ | -(-)--+-(-)(+) | - | -0.1 | -0.1 |
| *PVR* | chr19:45.2Mb | $6.1\times10^{-2}$ | **$3.0\times10^{-10}$** | $4.8\times10^{-2}$ | $6.3\times10^{-2}$ | **$1.1\times10^{-9}$** | $4.9\times10^{-2}$ | (-)++--+ | - | -0.1 | -0.1 |
| *LDLR* | chr19:11.2Mb | $1.8\times10^{-3}$ | $4.7\times10^{-5}$ | $3.8\times10^{-2}$ | $2.5\times10^{-1}$ | **$2.4\times10^{-7}$** | $5.2\times10^{-4}$ | +++++++++-++++--+ | - | - | 0.8 |
| | | | | | TG | | | | | | |
| *ANGPTL4* | chr19:8.4Mb | $2.6\times10^{-2}$ | **$1.2\times10^{-24}$** | $3.7\times10^{-2}$ | **$3.9\times10^{-25}$** | **$7.1\times10^{-24}$** | $2.6\times10^{-2}$ | (-)+---+--- | - | -0.3 | -0.2 |
| *LPL* | chr8:19.8Mb | $6.8\times10^{-1}$ | **$7.7\times10^{-20}$** | $2.6\times10^{-1}$ | **$1.8\times10^{-11}$** | **$4.6\times10^{-19}$** | $2.5\times10^{-2}$ | (+)+(+)--+- | - | 0.2 | 0.2 |

a. Gene position is defined based upon hg19, GRCh37 Genome Reference Consortium Human Reference 37

b. Direction of single site statistics for variants with MAF<5%. Variants within parenthesis have frequency >1%.

## Online Methods

This section starts with a summary of notation, and proceeds to describe the statistics to be shared between studies and single variant meta-analysis. We then show that the joint analysis statistics for different gene-level tests can be calculated using summary level data, enabling efficient meta-analysis. In the **Supplementary Material**, we provide many additional details and summarize how each of the test statistics used here can be derived as a score test using joint likelihood functions that allow for per-sample nuisance parameters.

### Notation

For simplicity, we describe our strategy for analysis of a single gene. Let $J$ be number of variant nucleotide sites genotyped in at least one study. For study $k$, let $N_k$ denote the number of samples phenotyped and genotyped, and let the vector $\vec{Y}_k = \left(Y_{1,k}, \cdots, Y_{N_k,k}\right)^T$ denote the quantitative trait residuals (after adjustment for any covariates), with variance $\sigma_k^2$. Within each study, we encode genotype information in matrix $\mathbf{X_k}$ where each entry $X_{i,j,k}$ represents the genotype for individual $i$ at site $j$, coded as the number of alternative alleles. We encode missing genotypes in the dataset as the average number of minor alleles in individuals who are genotyped for that marker. The multi-site genotype for individual $i$ is denoted by the row vector $\vec{X}_{i,\bullet,k}$, and the genotypes for all $N_k$ individuals at site $j$ are given by column vector $\vec{X}_{\bullet,j,k}$. For the ease of presentation, we define the mean genotype matrix $\overline{\mathbf{X}}_\mathbf{k}$, where the $(i,j)$-th element is $\left(\sum_i X_{i,j,k}\right)/N_k$.

### Summary Statistics To Be Shared

For each study, we first calculate and share a vector of score statistics $\vec{U}_k = \left(\mathbf{X_k} - \overline{\mathbf{X}}_\mathbf{k}\right)^T \vec{Y}_k$, a corresponding variance-covariance matrix $\mathbf{V_k} = \hat{\sigma}_k^2 N_k \operatorname{cov}(\mathbf{X_k}) = \hat{\sigma}_k^2 \left(\mathbf{X_k} - \overline{\mathbf{X}}_\mathbf{k}\right)^T \left(\mathbf{X_k} - \overline{\mathbf{X}}_\mathbf{k}\right)$, and allele frequencies for each marker $p_{j,k} = \sum_i X_{i,j,k} / 2N_k$ . Note that $\mathbf{V_k}$ effectively describes linkage disequilibrium relationships between the variants being examined. To perform quality control, we also share mean and variance for the quantitative trait residuals and genotype call rate and Hardy-Weinberg equilibrium p-values at each variant site.

**Meta-analysis of Single Variant Association Test Statistics**

We first combine single variant association test statistics across studies using the Cochran-Mantel-Haenszel method. Specifically, we calculate a score statistic at each site as:

$$T_{j,\bullet} = U_{j,\bullet} \Big/ \sqrt{V_{j,j,\bullet}}$$

where $U_{j,\bullet} = \sum_k U_{j,k}$ and $V_{j,j,\bullet} = \sum_k V_{j,j,k}$ . For ease of presentation, we denote the vector of single variant association tests after meta-analysis as $\vec{U} = \sum_k \vec{U}_k$ . Under the null, this vector is distributed as multivariate normal with mean vector $\vec{0}$ and covariance matrix $\sum_{\mathrm{k}} \mathbf{V_k}$ .

**Burden Tests That Assume Variants Have Similar Effect Sizes**

For a simple burden test in study *k*, the impact of multiple rare variants in a region can be modeled using a shared regression coefficient in a model that takes the form:

$$Y_{i,k} = \beta_{0,k} + \beta_{BURDEN} C_{BURDEN}\left(\vec{X}_{i,\bullet,k}\right) + \varepsilon_{i,k} \text{, where } \varepsilon_{i,k} \sim \mathrm{N}\left(0, \sigma_k^2\right)$$

$C_{BURDEN}\left(\vec{X}_{i,\bullet,k}\right)$ is a function that takes genotypes for a single individual as input and returns the count of rare alleles (the “rare variant burden”) in the gene being examined. When individual level data is available and nuisance parameters $\beta_{0,k}$ and $\sigma_k^2$ are allowed to vary between studies, the score statistic for a rare variant burden test becomes:

$$U_{BURDEN} = \sum_k U_{BURDEN,k} = \sum_k \vec{\omega}^T \vec{U}_k = \vec{\omega}^T \vec{U}$$

Under the null, this statistic is approximately normally distributed with mean 0 and variance $V_{BURDEN} = \vec{\omega}^T \left(\sum_k \mathbf{V_k}\right)\vec{\omega}$ , enabling significance tests. Here, $\vec{\omega}$ is the vector of weights, that is $\vec{\omega} = (\omega_1, \cdots, \omega_J)$, where $\omega_j$ is the weight assigned to variant *j* according to its allele frequency or its computationally predicted functional impact[10,14]. The formula above makes it clear that, when nuisance parameters are allowed to vary between studies, the same burden score statistics that could be calculated by sharing individual data can be equivalently calculated using shared summary statistics.

**Variable Threshold Tests with an Adaptive Frequency Threshold**

In variable threshold tests, rare variant burden statistics are calculated for each observed variant minor allele frequency threshold and significance is evaluated for the maximum of these statistics. Given a specific variant frequency threshold $F$ we define the resulting burden score statistic as:

$$U_{BURDEN(F)} = \vec{v}_F^T \vec{U} \ .$$

Here, $\vec{v}_F$ is a vector of indicators where the $j^{th}$ element equals 1 if the pooled minor allele frequency at variant site *j* is less than $F$ and zero otherwise. For convenience, we also define a matrix of minor allele frequency thresholds $\Phi = \left(\vec{v}_{F_1}, \vec{v}_{F_2}, \cdots, \vec{v}_{F_J}\right)$. After a burden statistic is calculated for each potential frequency threshold, these are standardized, dividing each statistic by its corresponding variance, and the maximum statistic is identified:

$$T_{VT} = \max_F \left\{T_{BURDEN(F)}\right\}, \text{ where } T_{BURDEN(F)} = U_{BURDEN(F)} \Big/ \sqrt{\vec{v}_F^T \sum_k \mathbf{V}_k \vec{v}_F} \ .$$

Significance for this statistic can be evaluated using the cumulative distribution function for the multivariate normal distribution[29]. Specifically, given the definition of the covariance between burden statistics calculated using different allele frequency thresholds, we have:

$$\left(T_{BURDEN(F_1)}, \cdots, T_{BURDEN(F_M)}\right) \sim \mathrm{MVN}\left(\vec{0}, \Phi^T \left(\sum_k \mathbf{V_k}\right)\Phi\right).$$

**Burden Tests that Assume a Distribution of Variant Effect Sizes (e.g. SKAT tests)**

The simple burden test and variable threshold test described above can be underpowered when variants with opposite phenotypic effects reside in the same gene and are grouped together, because the shared regression coefficient can average close to zero in that situation. To accommodate this setting, we consider an underlying distribution of rare variance effect sizes with mean zero and test whether the variance of this distribution $\tau^2$ is greater than zero.

When individual level data is available, association analysis in study *k* is performed using the following model

$$Y_{i,k} = \beta_{0,k} + \sum_j \beta_j X_{i,j,k} + \varepsilon_{i,k}, \text{ where } \varepsilon_{i,k} \sim \mathrm{N}\left(0, \sigma_k^2\right)$$

We make inferences about rare variant effect sizes $\vec{\beta} = (\beta_1, \beta_2, \cdots, \beta_J)$ by assuming these follow a common distribution with mean zero and variance $\tau^2$. Under the null, $\tau^2 = 0$. Following Wu et al[9], in **Supplementary Material** we derive the score statistic for this model and show that it can be calculated on the basis of per-study summary statistics:

$$Q = \left(\sum_k \vec{U}_k\right)^T \mathbf{K} \left(\sum_k \vec{U}_k\right)$$

Here, $\mathbf{K}$ is the kernel matrix that compares multi-site genotypes. A default choice[9] is a diagonal matrix $\mathbf{K} = diag(\omega_1, \omega_2, \cdots, \omega_J)$, with $\omega_j$ being the weight assigned to variant site $j$. The statistic $Q$ follows a mixture chi-square distribution[30], with mixture proportions given by the eigenvalues for the matrix $\left(\sum_k \mathbf{V_k}\right)^{1/2} \mathbf{K} \left(\sum_k \mathbf{V_k}\right)^{1/2}$.

**Monte-Carlo Method for Empirical Assessment of Significance**

The previous sections describe how a series of gene-level test statistics can be calculated and, for each one, propose a strategy for evaluating significance. In practice, evaluating the required numerical integrals can be challenging because variance-covariance matrices that are sometimes singular or nearly singular. Since single variant test statistics are distributed as:

$$\sum_k \vec{U}_k = \sum_k \vec{Y}_k^T \left(\mathbf{X_k} - \overline{\mathbf{X}}_\mathbf{k}\right) \sim \text{MVN}\left(\vec{0}, \sum_k \mathbf{V_k}\right)$$

To evaluate significance empirically, we sample random vectors from the distribution $\text{MVN}\left(\vec{0}, \sum_k \mathbf{V_k}\right)$ and calculate gene level rare variant test statistics for each of these sampled random vectors, allowing us to obtain an empirical distribution for any gene-level statistic[31]. As usual, p-values can then be evaluated by comparing the test statistics for the original data with those in this empirical distribution. For computational efficiency, we use an adaptive algorithm where a larger number of vectors are sampled when assessing small p-values and fewer vectors are sampled when assessing larger p-values[19].

**Conditional Analyses**

It is well known that, due to linkage disequilibrium, one or more common causal variants can result in shadow association signals at other nearby common variants. For common variants, Yang et al[25] have shown that linkage disequilibrium relationships between variants, estimated from external reference panels, can be used to enable conditional analysis in meta-analysis settings. For rare variants and gene-level tests, accurately describing relationships between variants is crucial and we recommend against the use of external reference panels. Instead, in the **Supplementary Material**, we describe how conditional analysis statistics can be derived for different gene-level test in our meta-analysis setting.

**Simulation of Population Genetic Data**

We simulated haplotypes using a coalescent model and the program *ms*[15]. We chose a demographic model consistent with European demographic history[4], including an ancestral bottleneck followed by more recent population differentiation and exponential growth. Model parameters were based upon estimates from large scale sequencing studies[32], as detailed in **Supplementary Material**.

### Meta-Analysis of Lipid Traits

Summary statistics were calculated for each participating study and shared to enable a central meta-analysis. In single variant and gene-base rare variant association analysis, age, age$^2$, sex and cohort specific covariates, such as principal components of ancestry were included in the analysis. Trait residuals were standardized using an inverse normal transformation. More detailed descriptions for each participating cohort are given in the **Supplemental Methods**.

# Supplementary Online Methods for

# Meta-Analysis of Gene Level Association Tests

Dajiang J. Liu et al.

**<u>Online Methods:</u>**

We describe a framework for meta-analysis of functional unit level rare variant association tests. The approach starts with meta-analysis of single variant association test statistics and then uses these to construct test statistics for genes or other functional units. We describe the implementation of several rare variant association tests and strategies for conditional analysis, which can provide a useful means of disentangling nearby signals. Finally, we propose a Monte Carlo simulation based strategy to aid in evaluating significance levels empirically. The document also includes a brief summary of the simulations carried out in preparing our manuscript.

**Notation**

We consider constructing joint analysis statistics of rare variant association tests using multiple studies. For simplicity, we describe our strategy for analysis of a single gene, but the approach naturally extends to multiple genes. Let $J$ be number of variant nucleotide sites of interest genotyped (using arrays or sequencing) in at least one of the studies. For study $k$ , let $N_k$ denote the number of samples phenotyped and genotyped, and let the vector $\vec{Y}_k = \left(Y_{1,k}, \cdots, Y_{N_k,k}\right)$ denote the quantitative trait (or quantitative trait residuals) each with variance $\sigma_k^2$ . In all analyses reported here, we applied an inverse normal transformation to trait residuals prior to analysis. In our preliminary analyses, this transformation reduced the impact of non-normally distributed phenotypes and led to better-behaved quantile-quantile plots.

Within each study, we encode genotype information in a matrix:

$$\mathbf{X_k} = \begin{pmatrix} X_{1,1,k} & \cdots & X_{1,j,k} & \cdots & X_{1,J,k} \\ \vdots & \ddots & \vdots & \ddots & \vdots \\ X_{i,1,k} & \cdots & X_{i,j,k} & \cdots & X_{i,J,k} \\ \vdots & \ddots & \vdots & \ddots & \vdots \\ X_{N_k,1,k} & \cdots & X_{N_k,j,k} & \cdots & X_{N_k,J,k} \end{pmatrix}$$

Each entry in matrix $X_{i,j,k}$ represents the genotype individual $i$ at site $j$, coded as the number of alternative alleles carried by the individual. We encode missing genotypes in the dataset as the average number of alternative alleles in individuals who are genotyped for that marker; alternatively, more advanced imputation algorithms (as implemented in MaCH[1], IMPUTE2[2] or BEAGLE[3]) could be used.

The multi-site genotype for individual $i$ is denoted by the row vector $\vec{X}_{i,\bullet,k}$, and the genotypes for all $N_k$ individuals at site $j$ are given by column vector $\vec{X}_{\bullet,j,k}$. For the ease of presentation, we define the mean genotype matrix $\overline{\mathbf{X}}_{\mathbf{k}}$, where the $(i, j)$-th element is $\left(\sum_i X_{i,j,k}\right)/N_k$ and the centered genotype matrix $\mathbf{X_k} - \overline{\mathbf{X}}_{\mathbf{k}}$.

**Summary Statistics**

For each study, we first calculate a vector of score statistics $\vec{U}_k = \left(\mathbf{X_k} - \overline{\mathbf{X}}_{\mathbf{k}}\right)^T \vec{Y}_k$ and a corresponding variance-covariance matrix $\mathbf{V_k} = \hat{\sigma}_k^2 N_k \operatorname{cov}(\mathbf{X_k}) = \hat{\sigma}_k^2 \left(\mathbf{X_k} - \overline{\mathbf{X}}_{\mathbf{k}}\right)^T \left(\mathbf{X_k} - \overline{\mathbf{X}}_{\mathbf{k}}\right)$. Then, to enable meta-analysis, we share the following summary statistics between studies:

a) Score statistics $\vec{U}_k$, which can be meta-analyzed across studies and then combined into gene-level statistics.

b) The covariance matrix for single variant score statistics $\mathbf{V_k}$. This variance-covariance matrix will later allow us to calculate the distribution of gene-level statistics that result from combining several single variant score statistics. In principle, sharing the full matrix would allow the most flexibility when grouping variants into genes during meta-analysis and for executing conditional analyses. In practice, we make two simplifications. First, because the matrix is symmetric, we share only its upper triangle. Second, because most gene level tests group nearby variants, we share only covariance information for markers <1 Mb apart.

c) Estimated alternative allele frequencies for each marker $p_{j,k} = \sum_i X_{i,j,k} / 2N_k$, which can be used to decide which variants to analyze based on frequency.

d) Mean and variance for the quantitative trait residuals, for debugging purposes and for quality control in multi-sample analyses. As usual, these are $\hat{\mu}_k = \sum_i Y_{i,k} / N_k$ and $\hat{\sigma}_k^2 = \sum_i (Y_{i,k} - \hat{\mu}_k)^2 / N_k$.

e) Genotype call rate and p-values for testing Hardy-Weinberg equilibrium at each variant site, for quality control and to aid in variant filtering.

**Meta-analysis of Single Variant Association Test Statistics**

We first combine single variant association test statistics across studies using the Cochran-Mantel-Haenszel method. Specifically, we calculate a score statistic at each site as:

$$T_{j,\bullet} = U_{j,\bullet} / \sqrt{V_{j,j,\bullet}}$$

where $U_{j,\bullet} = \sum_k U_{j,k}$ and $V_{j,j,\bullet} = \sum_k V_{j,j,k}$. Cochran-Mantel-Haenszel statistics deal gracefully with very rare variants because $U_{j,k}$ and $V_{j,j,k}$ remain defined (as zero) even when a variant is monomorphic or missing in a study. For ease of presentation, we denote the vector of single variant association tests after meta-analysis as $\vec{U} = \sum_k \vec{U}_k$. Under the null, this vector is distributed as multivariate normal $\vec{U} \sim \mathrm{MVN}(\vec{0}, \sum_k \mathbf{V_k})$.

## Meta-Analysis of Gene-level Rare-Variant Association Tests

We consider two major types of rare variant association methods: (i) burden tests that assume all variants in a gene influence the trait in the same direction, such as the test implemented in the GRANVIL test by Morris and Zeggini[4] and (ii) methods that allow variants with opposite effects to reside in the same gene, such as the variance component score test implemented in SKAT by Wu et al[5]. Below, we show that both types of method can be derived in a regression model, which allows adjusting for covariates. Furthermore, we illustrate how the corresponding gene level statistics can be derived from single site meta-analysis statistics and how the information stored in the variance-covariance matrix can be used to enable evaluating statistical significance.

## Burden Tests That Assume Variants Have Similar Effect Sizes

For a simple burden test in study *k*, the impact of multiple rare variants in a region can be modeled using a shared regression coefficient $\beta_{BURDEN}$ in a regression model that takes the form:

$$Y_{i,k} = \beta_{0,k} + \beta_{BURDEN} C_{BURDEN}\left(\vec{X}_{i,\bullet,k}\right) + \varepsilon_{i,k} \text{, where } \varepsilon_{i,k} \sim \mathrm{N}\left(0, \sigma_k^2\right)$$

$C_{BURDEN}\left(\vec{X}_{i,\bullet,k}\right)$ is a function that takes genotypes for a single individual as input and returns the rare variant burden for the gene being examined. Popular definitions for $C_{BURDEN}\left(\vec{X}_{i,\bullet,k}\right)$ include simple sum statistics and a weighted sum statistic $C_{BURDEN}\left(\vec{X}_{i,\bullet,k}\right) = \sum_j \omega_j X_{i,j,k}$, where $\omega_j$ is the weight assigned to variant *j* according to its allele frequency or its computationally predicted functional impact[6,7]. Note that in formulating this regression model, we allow the intercept $\beta_{0,k}$ and residual error $\sigma_k^2$ to vary between studies, but assume that $\beta_{BURDEN}$ is shared across studies. For convenience of notation, we define a vector of nuisance parameters $\vec{\theta}_k = (\beta_{0,k}, \sigma_k^2)$ and $\vec{\theta} = (\vec{\theta}_1, \cdots, \vec{\theta}_K)$, which are used in the likelihood function below.

As usual, the likelihood factors into a product of per study likelihoods:

$$L(\beta_{BURDEN}, \vec{\theta} | \vec{Y}, \mathbf{X}) = \prod_k L(\beta_{BURDEN}, \vec{\theta} | \vec{Y}_k, \mathbf{X_k})$$

In joint analysis with individual level data, the score statistic is thus a sum for per study score statistics:

$$\begin{aligned} U_{BURDEN} &= \frac{\partial \log L(\beta_{BURDEN}, \vec{\theta} | \vec{Y}, \mathbf{X})}{\partial \beta_{BURDEN}} \\ &= \sum_k \frac{\partial \log L(\beta_{BURDEN}, \vec{\theta}_k | \vec{Y}_k, \mathbf{X_k})}{\partial \beta_{BURDEN}} \\ &= \sum_k U_{BURDEN,k} \end{aligned}$$

Its variance can be derived using the Fisher information matrix, and following the derivations in Lin and Tang[7] (who studied a general framework for performing rare variant association tests), it can be shown that the variance for the score statistic in joint analysis of all individuals equals the sum of variances in each individual study $V_{BURDEN} = \sum_k V_{BURDEN,k}$ . Therefore, when nuisance parameters are allowed to vary between studies, a score test for joint analysis of individual level data (and allowing for study specific nuisance parameters) is equivalent to combining per study score statistics via the Cochran-Mantel-Haenszel method.

The arguments above indicate that the joint analysis statistic for gene-level test can be constructed when a per-study $U_{BURDEN,k}$ statistic is shared. But, because of the simple relationship between burden and single variant score statistics in each study (specifically, $U_{\beta_{BURDEN},k} = \vec{\omega}^T \vec{U}_k$), the joint analysis statistic for gene-level association tests can be calculated using shared single marker statistics. Specifically, the burden test score statistics becomes:

$$U_{BURDEN} = \sum_k U_{BURDEN,k} = \sum_k \vec{\omega}^T \vec{U}_k = \vec{\omega}^T \vec{U}$$

Under the null, this statistic is approximately normally distributed with mean 0 and variance $V_{BURDEN} = \vec{\omega}^T \left(\sum_k \mathbf{V_k}\right)\vec{\omega}$, enabling significance tests. Note that the regression coefficient $\beta_{BURDEN}$ can be interpreted as a weighted average of single variant effects[8].

**Variable Threshold Tests with an Adaptive Frequency Threshold**

In variable threshold tests, rare variant burden statistics are calculated for each potential definition of rare and significance is evaluated for the maximum of these statistics. Typically, to calculate these statistics all unique variant frequencies observed in a gene are listed and each of these frequencies is evaluated as a potential frequency threshold. Frequency thresholds can be defined in terms of the pooled minor allele frequency or, sometimes, the pooled minor allele count (the two can differ depending on whether samples where a variant is missing are assumed to be wild type or unknown).

Given a specific variant frequency threshold $F$ we define a corresponding burden score statistic as:

$$U_{BURDEN(F)} = \vec{v}_F^T \vec{U} \ .$$

Here, $\vec{v}_F$ is a vector of indicators where the $j^{th}$ element equals 1 if the pooled minor allele frequency at variant site $j$ is less than $F$ and zero otherwise. For convenience of presentation and without loss of generality, we also define a matrix of minor allele frequency thresholds $\Phi = \left(\vec{v}_{F_1}, \vec{v}_{F_2}, \cdots, \vec{v}_{F_J}\right)$. The covariance between burden score statistics $U_{BURDEN(\phi)}$ and $U_{BURDEN(\phi*)}$ calculated for thresholds $F$ and $F^*$, is equal to $\Omega_{BURDEN(F),BURDEN(F^*)} = \vec{v}_F^T \left(\sum_k \mathbf{V_k}\right) \vec{v}_{F*}$. After burden statistics are calculated for each potential frequency threshold, they are standardized, dividing each statistic by its corresponding variance, and the maximum statistic is identified:

$$T_{VT} = \max_F \{T_{BURDEN(F)}\}, \text{ where } T_{BURDEN(F)} = U_{BURDEN(F)} \Big/ \sqrt{\Omega_{BURDEN(F),BURDEN(F)}} .$$

Significance for this statistic can be evaluated using the cumulative distribution function for multivariate normal distribution. Specifically, given the definition of the covariance between burden statistics calculated using different allele frequency thresholds, we have:

$$\left(T_{BURDEN(F_1)}, \cdots, T_{BURDEN(F_M)}\right) \sim \mathrm{MVN}\left(\vec{0}, \Phi^T \left(\sum_k \mathbf{V_k}\right) \Phi\right).$$

Significance tests can be calculated using standard methods for calculating multivariate normal integrals[9]:

$$\begin{aligned} &\Pr\left(T_{VT} < t \mid \left(T_{BURDEN(F_1)}, \cdots, T_{BURDEN(F_J)}\right) \sim \mathrm{MVN}\left(\vec{0}, \Phi^T \mathbf{V} \Phi\right)\right) = \\ &\Pr\left(T_{BURDEN(F_1)} < t, \cdots, T_{BURDEN(F_J)} < t \mid \left(T_{BURDEN(F_1)}, \cdots, T_{BURDEN(F_J)}\right) \sim \mathrm{MVN}\left(\vec{0}, \Phi^T \mathbf{V} \Phi\right)\right) \end{aligned}$$

In practice, the covariance matrix for burden score statistics calculated using different frequency thresholds can be singular or nearly so and evaluating the corresponding integrals can be numerically challenging. We recommend verifying analytical p-values using simulations (described later in this document).

**Gene-level Tests that Assume a Distribution of Variant Effect Sizes (e.g. SKAT tests)**

The simple burden test and variable threshold test described above can be underpowered when variants with opposite effects on the phenotype reside in the same gene and are grouped together, because the shared regression coefficient can average close to zero in that situation. One option in this setting would be to model the effects of each rare variant individually – but that strategy consumes many degrees of freedom and thus loses efficiency. Instead, we assume an underlying distribution of rare variance effect sizes with mean zero and test whether the variance of this distribution, $\tau^2$, is greater than zero.

Specifically, we consider the model:

$$Y_{i,k} = \beta_{0,k} + \sum_j \beta_j X_{i,j,k} + \varepsilon_{i,k}, \text{ where } \varepsilon_{i,k} \sim \mathrm{N}\left(0, \sigma_k^2\right),$$

and make inferences about rare variant effect sizes $\vec{\beta} = (\beta_1, \beta_2, \cdots, \beta_J)$ by assuming these follow a common distribution with mean zero and variance $\tau^2$. Under the null, $\vec{\beta} = 0$, which is equivalent to $\tau^2 = 0$. Following Wu et al[5], we consider the likelihood:

$$\begin{aligned} L\left(\tau, \vec{\theta} \middle| \vec{Y}, \mathbf{X}\right) &= \int L\left(\vec{\beta}, \vec{\theta} \middle| \vec{Y}, \mathbf{X}\right) p\left(\vec{\beta} \middle| \tau\right) d\vec{\beta} \\ &= \int \prod_k L\left(\vec{\beta}, \vec{\theta}_k \middle| \vec{Y}_k, \mathbf{X_k}\right) p\left(\vec{\beta} \middle| \tau\right) d\vec{\beta} \\ &= \int \exp\left(\sum_k \log\left(L\left(\vec{\beta}, \vec{\theta}_k \middle| \vec{Y}_k, \mathbf{X_k}\right)\right)\right) p\left(\vec{\beta} \middle| \tau\right) d\vec{\beta} \end{aligned}$$

In order to derive the variance component score statistics for this likelihood, we apply a Laplace transformation to the marginal likelihood function, as suggested by Lin[10]. If repeating this calculation, please note that the integrant satisfies:

$$\exp\left(\log L\left(\vec{\beta},\hat{\vec{\theta}}_k\middle|\vec{Y}_k,\mathbf{X_k}\right)\right)=\exp\left(\log L\left(0,\hat{\vec{\theta}}_k\middle|\vec{Y}_k,\mathbf{X_k}\right)\right)\times$$

$$\left(1+tr\left(\vec{\beta}^T\left(\left(\frac{\partial\log L\left(\vec{\beta},\hat{\vec{\theta}}_k\middle|\vec{Y}_k,\mathbf{X_k}\right)}{\partial\vec{\beta}}\right)^T\frac{\partial\log L\left(\vec{\beta},\hat{\vec{\theta}}_k\middle|\vec{Y}_k,\mathbf{X_k}\right)}{\partial\vec{\beta}}+\frac{\partial^2\log L\left(\vec{\beta},\hat{\vec{\theta}}_k\middle|\vec{Y}_k,\mathbf{X_k}\right)}{\partial\vec{\beta}^2}\right)\vec{\beta}\right)+o\left(\left|\vec{\beta}\right|^2\right)\right)$$

Then, following the argument in Lin[10], it can be shown that the variance component score statistics in the joint analysis with individual level data for testing $\tau=0$ is:

$$\begin{aligned}Q&=\left(\sum_k\frac{\partial\log L\left(\vec{\beta},\hat{\vec{\theta}}_k\middle|\vec{Y}_k,\mathbf{X_k}\right)}{\partial\vec{\beta}}\right)^T\mathbf{K}\left(\sum_k\frac{\partial\log L\left(\vec{\beta},\hat{\vec{\theta}}_k\middle|\vec{Y}_k,\mathbf{X_k}\right)}{\partial\vec{\beta}}\right)\\&=\left(\sum_k\vec{U}_k\right)^T\left(\sum_k\vec{U}_k\right)\end{aligned}$$

Therefore, gene level statistics that allow for a distribution of rare variant effect sizes (rather than modeling these using a shared regression coefficient) can also be constructed after meta-analysis of single variant statistics. Note that $Q$ is a quadratic function of the single site meta-analysis statistics, in contrast to burden statistics defined in previous sections, which are linear functions of single site statistics. In practice, weights can also be assigned in the variance component score statistics, and the test statistic takes the form of $Q=\left(\sum_k\vec{U}_k\right)^T\mathbf{K}\left(\sum_k\vec{U}_k\right)$. The matrix $\mathbf{K}$ is the kernel that compares multi-site genotypes. A default choice is $\mathbf{K}=diag(\omega_1,\omega_2,\cdots,\omega_J)$, with $\omega_j$ being the weight assigned to variant site $j$ [5]. The statistic $Q$ follows a mixture chi-square distribution[11], with mixture proportions given by the eigenvalues for the matrix $\left(\sum_k\mathbf{V_k}\right)^{1/2}\mathbf{K}\left(\sum_k\mathbf{V_k}\right)^{1/2}$.

**Monte Carlo Method for Empirical Assessment of Significance**

The previous sections describe how a series of gene-level test statistics can be calculated and, for each one, propose a strategy for evaluating significance. In practice, evaluating the required numerical integrals can be challenging, because multiple variants or a set of burden scores can produce co-linear sets of predictors and variance-covariance matrices that are singular or nearly so. In this section, we describe a simple strategy for re-sampling plausible sets of single marker test statistics. Gene level statistics can then be evaluated for each of these simulated vectors of single marker test statistics to assess significance empirically, in order to avoid some of the problems inherent with numerical integration.

Recall that test statistics are distributed as:

$$\sum_k \vec{U}_k = \sum_k \vec{Y}_k^T \left(\mathbf{X_k} - \overline{\mathbf{X}}_\mathbf{k}\right) \sim \mathrm{MVN}\left(\vec{0}, \sum_k \mathbf{V_k}\right)$$

To evaluate significance empirically, we sample random vectors from the distribution $\mathrm{MVN}\left(\vec{0}, \sum_k \mathbf{V_k}\right)$ and calculate gene level rare variant test statistics for each of these sampled random vectors, allowing us to obtain an empirical distribution for any gene-level statistic[12]. As usual, p-values can then be evaluated by comparing the test statistics for the original data with those in this empirical distribution. For computational efficiency, we use an adaptive algorithm where a larger number of vectors are sampled when assessing small p-values and fewer vectors are sampled when assessing larger p-values[13]. Specifically, we continue sampling new vectors until the number of sampled statistics greater than the statistic in the original data exceeds a particular threshold (100, unless noted otherwise) or the total number of sampled vectors exceeds a predefined limit (40,000,000; unless noted otherwise).

**Conditional Analyses**

It is well known that, due to linkage disequilibrium, one or more common causal variants can result in shadow association signals at other nearby common variants. As illustrated in our analysis of the *APOE* locus in the text, common variant association signals can also result in shadow rare variant association signals at nearby genes. Conditional analysis provides a useful procedure for disentangling neighboring association signals in this setting, for example by checking whether weaker signals remain significant after conditioning on nearby strong signals.

For common variants, Yang et al[14] have shown that linkage disequilibrium relationships between variants, estimated from external reference panels, can be used to enable conditional analysis in meta-analysis settings. For rare variants and gene-level tests, accurately describing relationships between variants is crucial and we recommend against the use of external reference panels. Instead, we recommend using linkage disequilibrium relationships estimated in the samples being analyzed and summarized in the variance-covariance matrix of single variant score statistics.

To describe our strategy for conditional analyses, we first decompose the genotype matrix into two components: a matrix of genotypes $\mathbf{X_k}$ for variants to be tested for independent association and a matrix of genotypes $\mathbf{Z_k}$ for variants that should be included as covariates in the null regression model (and, thus, controlled for). In order to facilitate presentations, we denote $\mathbf{W_k} = (\mathbf{X_k}, \mathbf{Z_k})$.

**Conditional Analysis for Gene Level Tests That Use A Shared Regression Coefficient for Rare Variants**

When individual level data is available, conditional analysis considers a model similar to:

$$Y_{i,k} = \beta_{0,k} + \beta_{BURDEN} C_{BURDEN}\left(\vec{X}_{i,\bullet,k}\right) + \vec{\alpha}_k^T \vec{Z}_{i,k} + \varepsilon_{i,k}$$

This analysis could be readily carried out by repeating analysis and calculation of score statistics for each study, but this is not required. Instead, the score statistics that result from the conditional analysis described above can be readily estimated using summary information.

Let $\tilde{T}_{\beta_{BURDEN},k}, \tilde{U}_{\beta_{BURDEN},k}$ and $\tilde{V}_{\beta_{BURDEN},\beta_{BURDEN},k}$ denote test statistics, score statistics and their variances from conditional analysis; analogous to statistics previously defined for unconditional analysis. As usual, $\tilde{T}_{\beta_{BURDEN},k} = \tilde{U}_{\beta_{BURDEN},k} \Big/ \sqrt{\tilde{V}_{\beta_{BURDEN},\beta_{BURDEN},k}}$ . To derive the component statistics, we use the approach of Lin and Tang[7], to show that:

$$\tilde{U}_{\beta_{BURDEN},k} = \left( \left( \vec{Y}_k - \bar{Y}_k \right) - \left( \mathbf{Z_k} - \bar{\mathbf{Z}}_\mathbf{k} \right) \hat{\vec{\alpha}}_k \right)^T \mathbf{X_k}$$

$$\text{with } \hat{\alpha}_k = \left( \left( \mathbf{Z_k} - \bar{\mathbf{Z}}_\mathbf{k} \right)^\mathbf{T} \left( \mathbf{Z_k} - \bar{\mathbf{Z}}_\mathbf{k} \right) \right)^{-1} \left( \mathbf{Z_k} - \bar{\mathbf{Z}}_\mathbf{k} \right)^\mathbf{T} \vec{Y}_k ,$$

and that:

$$\begin{aligned} \tilde{V}_{\beta_{BURDEN},\beta_{BURDEN},k} = {} & \hat{\varphi}^2 \omega^T \left( \left( \mathbf{X_k} - \bar{\mathbf{X}}_\mathbf{k} \right)^\mathbf{T} \left( \mathbf{X_k} - \bar{\mathbf{X}}_\mathbf{k} \right) \right) \omega \\ & - \hat{\varphi}^2 \omega^T \left( \left( \mathbf{X_k} - \bar{\mathbf{X}}_\mathbf{k} \right)^\mathbf{T} \left( \mathbf{Z_k} - \bar{\mathbf{Z}}_\mathbf{k} \right) \right) \left( \left( \mathbf{Z_k} - \bar{\mathbf{Z}}_\mathbf{k} \right)^\mathbf{T} \left( \mathbf{Z_k} - \bar{\mathbf{Z}}_\mathbf{k} \right) \right)^{-1} \left( \left( \mathbf{Z_k} - \bar{\mathbf{Z}}_\mathbf{k} \right)^\mathbf{T} \left( \mathbf{X_k} - \bar{\mathbf{X}}_\mathbf{k} \right) \right) \omega \end{aligned} ,$$

$$\text{with } \hat{\varphi}^2 = 1/N_k \times \left( \left( \vec{Y}_k - \bar{Y}_k \right) - \left( \mathbf{Z_k} - \bar{\mathbf{Z}}_\mathbf{k} \right) \hat{\vec{\alpha}}_k \right)^T \left( \left( \vec{Y}_k - \bar{Y}_k \right) - \left( \mathbf{Z_k} - \bar{\mathbf{Z}}_\mathbf{k} \right) \hat{\vec{\alpha}}_k \right)$$

Now, we can verify that $\tilde{U}_{\beta_{BURDEN},k}$ and $\tilde{V}_{\beta_{BURDEN},\beta_{BURDEN},k}$ can be calculated using shared summary level statistics, because all key terms in the above equations can be extracted from the list of single variant score statistics and from the variance-covariance matrix of single marker association test statistics (which we have shared), since

$$\left( \mathbf{W_k} - \bar{\mathbf{W}}_\mathbf{k} \right)^\mathbf{T} \left( \mathbf{W_k} - \bar{\mathbf{W}}_\mathbf{k} \right) = \begin{pmatrix} \left( \mathbf{X_k} - \bar{\mathbf{X}}_\mathbf{k} \right)^\mathbf{T} \left( \mathbf{X_k} - \bar{\mathbf{X}}_\mathbf{k} \right) & \left( \mathbf{X_k} - \bar{\mathbf{X}}_\mathbf{k} \right)^\mathbf{T} \left( \mathbf{Z_k} - \bar{\mathbf{Z}}_\mathbf{k} \right) \\ \left( \mathbf{Z_k} - \bar{\mathbf{Z}}_\mathbf{k} \right)^\mathbf{T} \left( \mathbf{X_k} - \bar{\mathbf{X}}_\mathbf{k} \right) & \left( \mathbf{Z_k} - \bar{\mathbf{Z}}_\mathbf{k} \right)^\mathbf{T} \left( \mathbf{Z_k} - \bar{\mathbf{Z}}_\mathbf{k} \right) \end{pmatrix} .$$

Finally, meta-analysis burden score statistics can be calculated as:

$$\tilde{T}_{\beta_{BURDEN},\bullet} = \sum_k \tilde{U}_{\beta_{BURDEN},k} \Big/ \sqrt{\sum_k \tilde{V}_{\beta_{BURDEN},\beta_{BURDEN},k}} \ .$$

Thus, conditional statistics after meta-analysis can be calculated from shared single variant statistics and their variance-covariance matrix, as desired.

**Conditional Analysis for Tests that Assume a Distribution of Rare Variant Effect Sizes (e.g. SKAT)**

Similar arguments can be used to derive formulae for conditional analysis of rare variant association tests in settings where direction of effect and effect sizes are allowed to vary between markers. In that setting, we follow the approach of Wu et al[5]. The variance component score test takes the form:

$$\tilde{Q}_k = \left(\left(\vec{Y}_k - \bar{Y}_k\right) - \left(\mathbf{Z_k} - \bar{\mathbf{Z}}_\mathbf{k}\right)\hat{\vec{\alpha}}_k\right)^T \mathbf{X_k} \times \mathbf{K} \times \mathbf{X_k^T}\left(\left(\vec{Y}_k - \bar{Y}_k\right) - \left(\mathbf{Z_k} - \bar{\mathbf{Z}}_\mathbf{k}\right)\hat{\vec{\alpha}}_k\right)$$

Following the derivation for unconditional analysis, the meta-analysis test statistic is given by :

$$\tilde{Q} = \left(\sum_k \left(\left(\vec{Y}_k - \bar{Y}_k\right) - \left(\mathbf{Z_k} - \bar{\mathbf{Z}}_\mathbf{k}\right)\hat{\vec{\alpha}}_k\right)^T \mathbf{X_k}\right) \times \mathbf{K} \times \left(\sum_k \mathbf{X_k^T}\left(\left(\vec{Y}_k - \bar{Y}_k\right) - \left(\mathbf{Z_k} - \bar{\mathbf{Z}}_\mathbf{k}\right)\hat{\vec{\alpha}}_k\right)\right)$$

Then, noting that the single variant score statistics $\sum_k \left(\left(\vec{Y}_k - \bar{Y}_k\right) - \left(\mathbf{Z_k} - \bar{\mathbf{Z}}_\mathbf{k}\right)\hat{\vec{\alpha}}_k\right)^T \mathbf{X_k}$ follow a multivariate normal distribution with mean zero and variance-covariance matrix $\sum_k \tilde{\mathbf{V}}_k$ , where

$$\begin{aligned}\tilde{\mathbf{V}}_k &= \hat{\varphi}^2\left(\left(\mathbf{X_k} - \bar{\mathbf{X}}_\mathbf{k}\right)^\mathbf{T}\left(\mathbf{X_k} - \bar{\mathbf{X}}_\mathbf{k}\right)\right) \\ &\quad - \hat{\varphi}^2\left(\left(\mathbf{X_k} - \bar{\mathbf{X}}_\mathbf{k}\right)^\mathbf{T}\left(\mathbf{Z_k} - \bar{\mathbf{Z}}_\mathbf{k}\right)\right)\left(\left(\mathbf{Z_k} - \bar{\mathbf{Z}}_\mathbf{k}\right)^\mathbf{T}\left(\mathbf{Z_k} - \bar{\mathbf{Z}}_\mathbf{k}\right)\right)^{-1}\left(\left(\mathbf{Z_k} - \bar{\mathbf{Z}}_\mathbf{k}\right)^\mathbf{T}\left(\mathbf{X_k} - \bar{\mathbf{X}}_\mathbf{k}\right)\right)\end{aligned}$$

It is straightforward to show that $\tilde{Q}$ follows a mixture chi-square distribution with mixture proportions being the eigenvalues of $\left(\sum_\mathbf{k} \tilde{\mathbf{V}}_\mathbf{k}\right)^{1/2} \mathbf{K} \left(\sum_\mathbf{k} \tilde{\mathbf{V}}_\mathbf{k}\right)^{1/2}$. Therefore, score statistics and the variance-covariance matrix for single marker statistics, are sufficient to enable derivation of statistics and p-values for conditional meta-analysis.

**Simulation of Population Genetic Data**

We simulated haplotypes using a coalescent model and the program *ms*[15]. We chose a demographic model consistent with European demographic history[16], including an ancestral bottleneck followed by more recent population differentiation and exponential growth (**Supplemental Figure 1**). Model parameters were based upon estimates from large scale sequencing studies[17], tuned such that measures of genetic diversity between simulated sub-populations match estimates from European samples[18]. The simulated haplotypes had an average pairwise sequence difference of $\pi = 0.001$ and an average $F_{ST} = 0.004$. Furthermore, when 5000 haplotypes were sampled, a typical simulated 5000 base pair region included ~100 variant sites, of which 80% had MAF<1% and ~49% were singletons. For any two pools of 5000 haplotypes sampled from different subpopulations, ~3% of variants with MAF<1% and 35% of the variants with MAF>1% are shared, consistent with expectations from population genetics and observations from real data[17,19].

Our model assumes an ancestral population with effective population size of $N_1 = 10{,}000$ where an instantaneous bottleneck event 3,000 generations in the past reduced population size to $N_{bottleneck}$ = 75. Then, our simulations assume that this population simultaneously split into present day populations 500 generations before the present. Following the divergence from the ancestral population, the present-day populations underwent recent exponential growth, each growing to a present day effective population size of $N_0 = 1\times10^6$ over 400 generations. We assume equal, symmetric migration rates between the sub-populations with a per-haplotype, per-generation migration rate of $5\times10^{-4}$. In the simulation, a per-basepair, per-generation mutation rate of $2.5\times10^{-8}$ and a recombination rate equivalent to 1cM/Mb were assumed.

**Type I Error Rate and Power**

Using simulated genetic data, we estimated power and type I error for each of the methods described here. We considered three representative rare variant tests association tests: a simple burden test (with MAF thresholds of 1% and 5%), a variable threshold association test, and the SKAT test. First, we generated 50,000,000 null replicates to evaluate type I error rates in meta-analyses of 3, 6 and 9 samples of 1000 individuals at significance level $\alpha$=0.001, 0.0001, or $2.5\times10^{-6}$. As shown in **Supplemental Table 1**, the type I error rates are well controlled.

Next, the power for different rare variant tests in meta-analysis was evaluated for a series of genetic models. In the first set of simulations, 20 or 50% of variants were randomly chosen as causal. In a second set of simulations, only variants with MAF<0.5% have potential phenotypic effects and 20 or 50% of these variants with MAF<0.5% were selected to be causal at random. In both cases, we considered situations where: (i) all causal variants increase expected trait values by 0.25 standard deviations; (ii) 80% of causal variants increase trait values by 0.25 standard deviation units, while the remaining 20% decrease trait values; (iii) effect sizes for each variant, measured in standard deviation unit, were sampled from a normal distribution with mean 0 and variance 0.25.

**Supplemental Figures 4 and 5** summarize the results of our power simulations, considering the various simulation scenarios and meta-analysis of up to 100 samples of 1000 individuals each. Several patterns are clear from the figures. First, for the effect sizes simulated here, very large sample sizes may be required to ensure adequate power. In some settings, power only reaches ~60% in analyses that include ~100,000 individuals, even using the most powerful available test. Second, we did not find a universally most powerful method, emphasizing the utility of approaches such as ours that can be extended to implement a diverse set of test statistics. Typically, we find that when the proportion of non-causal variants is high or causal variants can have opposite effects, the SKAT was more powerful. When causal variants have effects in the same direction, simpler burden tests were more powerful.

**Supplemental Figure 6** shows that, as expected, pooled analysis of individual level data and meta-analysis of summary statistics, as proposed here, result in nearly identical power.

**Evaluation of Conditional Analysis Strategy**

As described in our analysis of genes neighboring *APOE*, common variant association signals can produce inflated rare variant test statistics at nearby genes due to linkage disequilibrium. To evaluate our strategy for conditional analysis of rare variant association tests, we selected one common variant with pooled MAF>10% as causal and increase the mean trait value by 0.25 standard deviation. We then evaluated the type I error rate of gene-level rare variant association test statistics (**Supplemental Figure 8**). The results show that, without conditioning, p-values deviate substantially from null expectations. The results also show that, after conditioning, p-values for rare variant association tests behave as expected under the null.

**Meta-Analysis of Lipid Traits**

Summary statistics were calculated for each participating study and shared to enable a central meta-analysis. In single variant and gene-base rare variant association analysis, age, $age^2$, sex and cohort specific covariates, such as principal components of ancestry were included in the analysis. Trait residuals were standardized using an inverse normal transformation.

**STUDY DESCRIPTIONS**

**Malmö Diet and Cancer Study – Cardiovascular Cohort (MDC-CC)**

The Malmö Diet and Cancer Study[20] is a community-based prospective epidemiologic cohort of 28,449 persons recruited for a baseline examination between 1991 and 1996. From this cohort, 6,103 persons were randomly selected to participate in the cardiovascular cohort, which sought to investigate risk factors for cardiovascular disease. All participants underwent a medical history assessment and a physical examination.

**Women's Health Initiative**

The WHI[21] encompasses four randomized clinical trials as well as a prospective cohort study of 161,808 post-menopausal women aged 50–79, recruited (1993–1998) and followed up at 40 centers across the US. Samples examined here were genotyped as part of the NHLBI Exome Sequencing Project.

**Ottawa Heart Study**

Cases and controls were recruited from either the lipid clinic at the University of Ottawa Heart Institute or the cardiac catheterization laboratory[22]. All cases were required to have at least one of: a stenosis in a major epicardial vessel of at least 50%; have had a percutaneous intervention (PCI); have had coronary artery bypass surgery (CABG); or have had a myocardial infarction (MI). Cases with diabetes mellitus were excluded. Age of onset of CAD was required to be ≤55 years old for men and ≤65 years old for women. Controls were either healthy asymptomatic elderly individuals or were recruited through the catheterization laboratory with no stenosis ≥50% in any major epicardial vessel and were required to be ≥65 years old for men and ≥70 years old for women.

**PROCARDIS**

The PROCARDIS[23] "genetically-enriched" case collection is composed of sibships (proband and at least one affected sibling) with coronary disease. Ascertainment criteria for PROCARDIS probands were myocardial infarction (MI) or symptomatic acute coronary syndrome before the age of 66 years. For each of the coronary disease cases included in the "genetically-enriched" case-control study, it was planned to recruit one control of the same sex, ethnicity and within 5 years of age of cases, with no personal or sibling history of coronary disease before age 66 years. In the UK, controls were identified by mailing a self-administered questionnaire to spouses or siblings of spouses or male friends of any individuals who had previously returned a completed questionnaire to the PROCARDIS study. Eligible respondents were asked to attend their general practice to have their blood pressure, height and weight recorded, and to provide a blood sample. In Sweden, Italy and Germany, controls were selected from population registers and invited to attend a special clinic to have their blood pressure, height and weight recorded, to provide a blood sample and to complete a self-administered questionnaire.

**HUNT – The Nord-Trøndelag Health Study**

The HUNT study has been described in detail previously[24]. The HUNT study is a population based health study with personal and family medical histories on 106,436 people from Nord-Trøndelag County, Norway, collected during three phases from 1984 to 2008. A subsample of 5,869 individuals were successfully genotyped on the iSelect Exomchip V1.0 (Illumina, San Diego, CA), 2,928 cases with retrospectively hospital diagnosed myocardial infarction and 2,941 healthy controls matched on sex, birth year and municipality. Genotype calling was done using GenTrain version 2.0 in GenomeStudio V2011.1 (Illumina, San Diego, CA) in combination with zCall version 2.2 [25].

**Supplemental Figure 1**: Demographic model for simulated European populations. The demographic model includes an ancient population bottleneck, recent exponential growth, differentiation and migration. The model parameters were calibrated to mimic populations sampled in continental Europe.

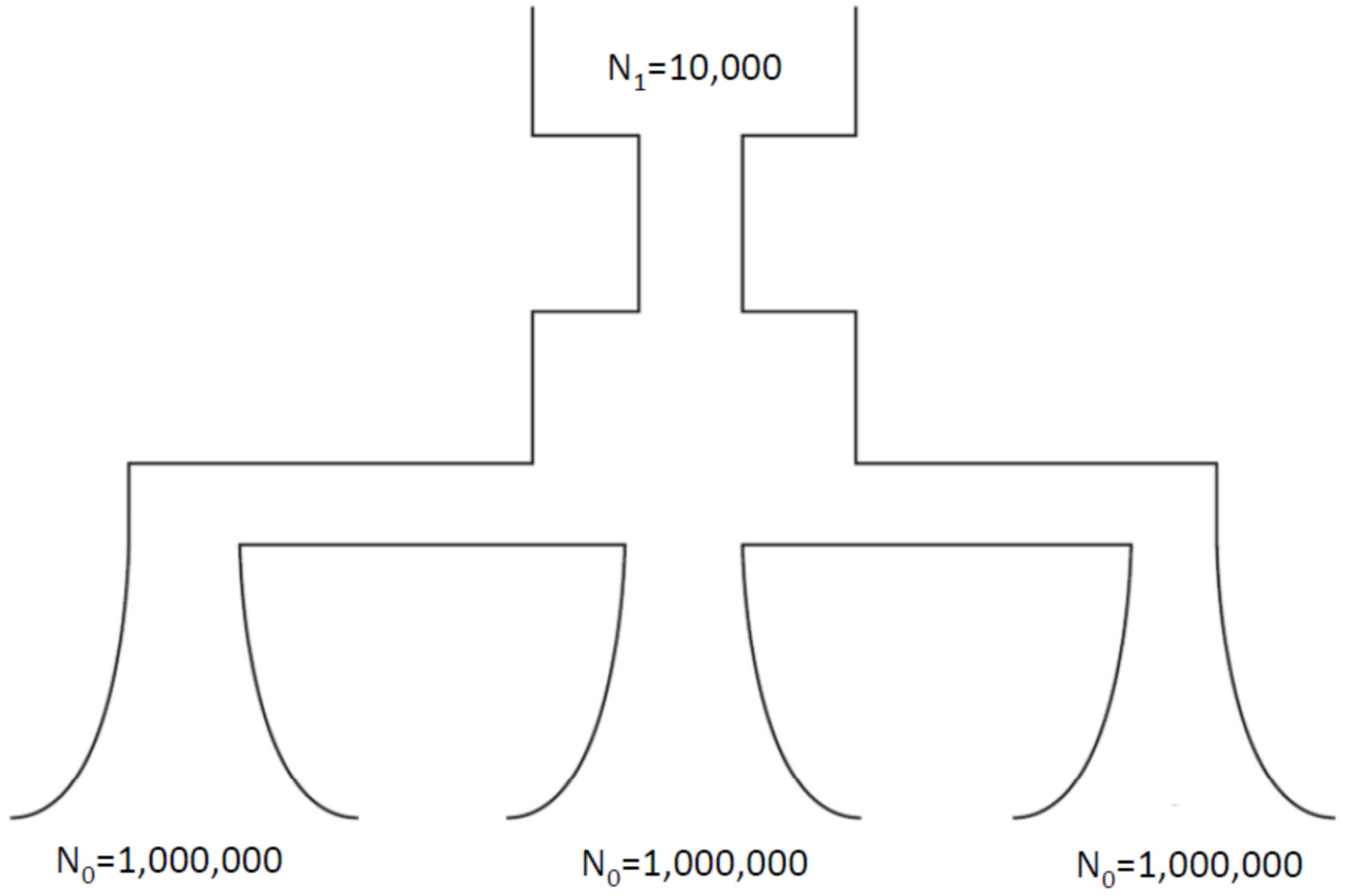

**Supplemental Figure 2**: Comparison of statistics and p-values for a simple burden test, variable threshold (VT) and sequence kernel association test (SKAT) in analysis of pooled samples (X-axis) and in meta-analysis (Y-axis). Three samples of 1000 individuals were simulated. Traits were generated assuming that 50% of the variants are causal and that, among these, 80% of the variants increase the trait values by 0.25 standard deviation units and the remaining 20% decrease trait values by the same amount. Empirical p-values for the variable threshold (VT) test were obtained using the adaptive Monte Carlo procedure, stopping after 100 simulated statistics were greater than the original statistic or the number of simulations exceeded 40,000,000.

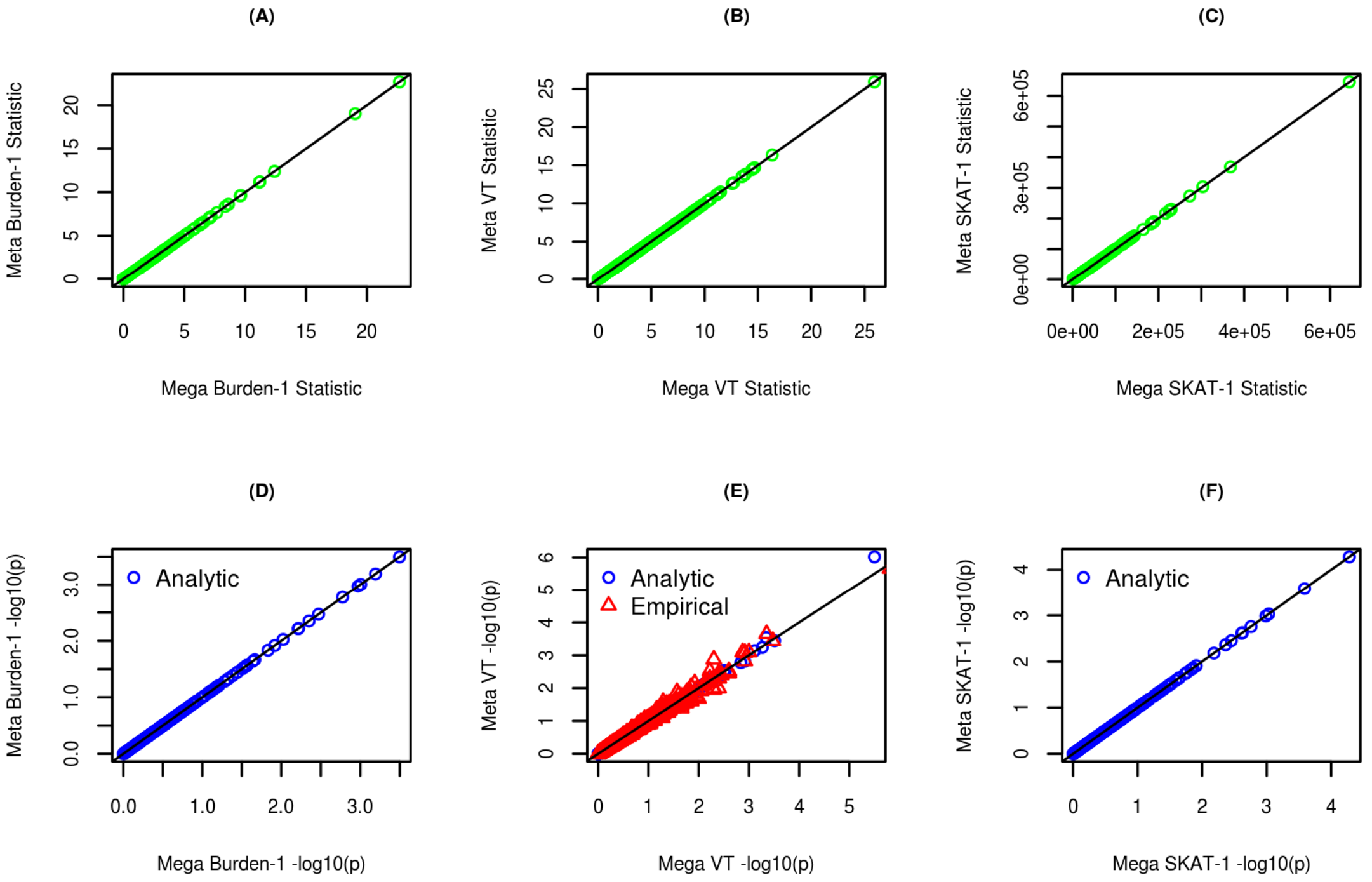

**Supplemental Figure 3**: Comparison of statistics and p-values for a simple burden test, variable threshold (VT) and sequence kernel association test (SKAT) in analysis of pooled samples (X-axis) and in meta-analysis (Y-axis). This Figure is analogous to **Supplementary Figure** 2, but assumes a random effect for each causal variant, distributed as Normal(0.25, 0.01) in standard deviation units.

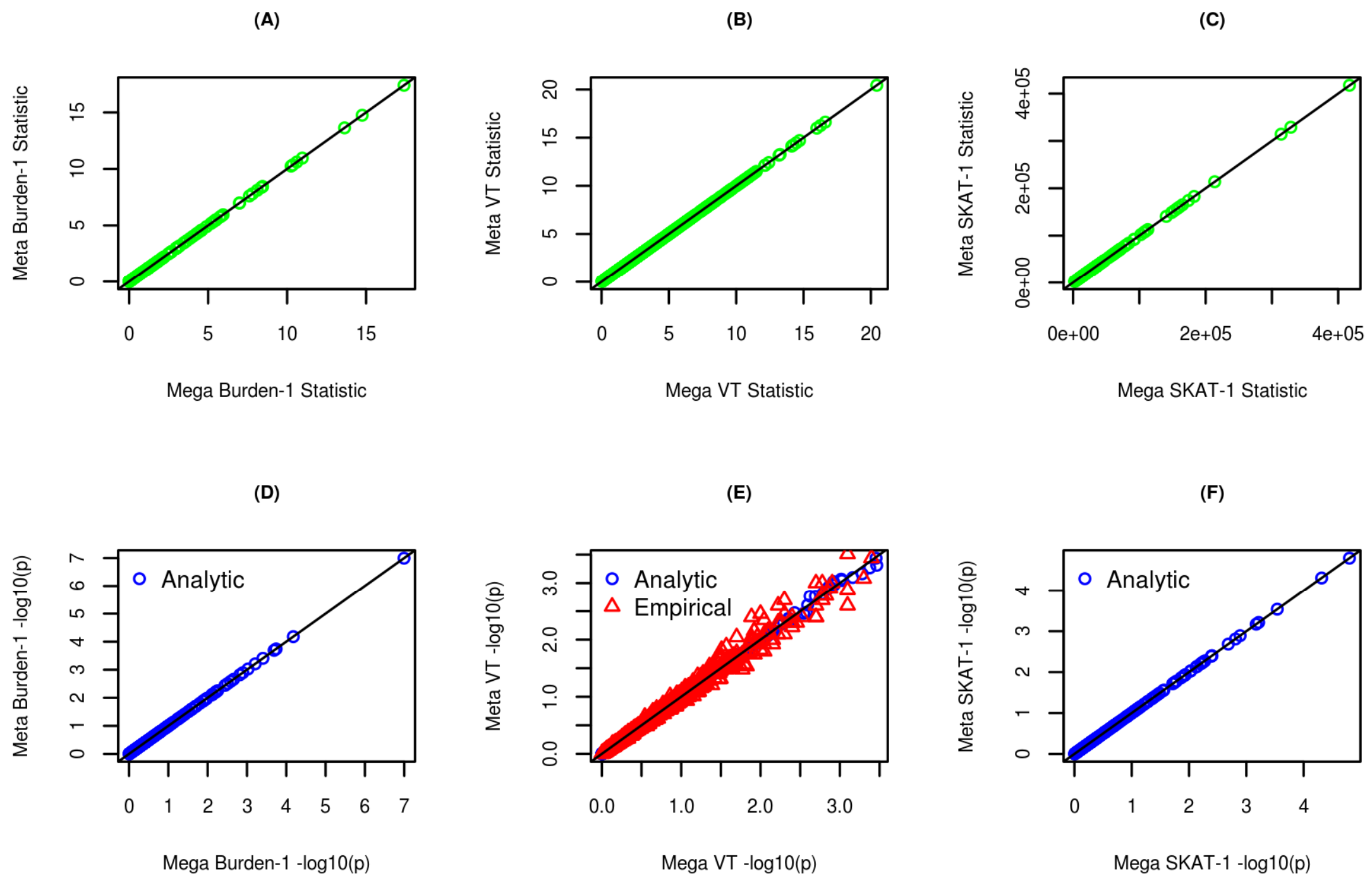

**Supplemental Figure 4:** Power of meta-analysis rare variant association tests as a function of the number of studies in the meta-analysis, each with 1000 individuals. Quantitative trait data were simulated assuming the trait was influenced by rare variants with frequency less than 5%. In panel A, 20% of the variants with MAF <5% are causal and each increases trait values by 0.25 standard deviation units. In panel B, 50% of the variants with MAF <5% are causal and each increases trait values by 0.25 standard deviation units. In panel C, 20% of the variants with MAF <5% are causal, with 80% of these increasing the trait by 0.25 standard deviation units with the other 20% decreasing trait values by the same amount. In panel D, 50% of the variants with MAF <5% are causal, with 80% of these increasing the trait by 0.25 standard deviation units while the other 20% decreasing trait values by the same amount.

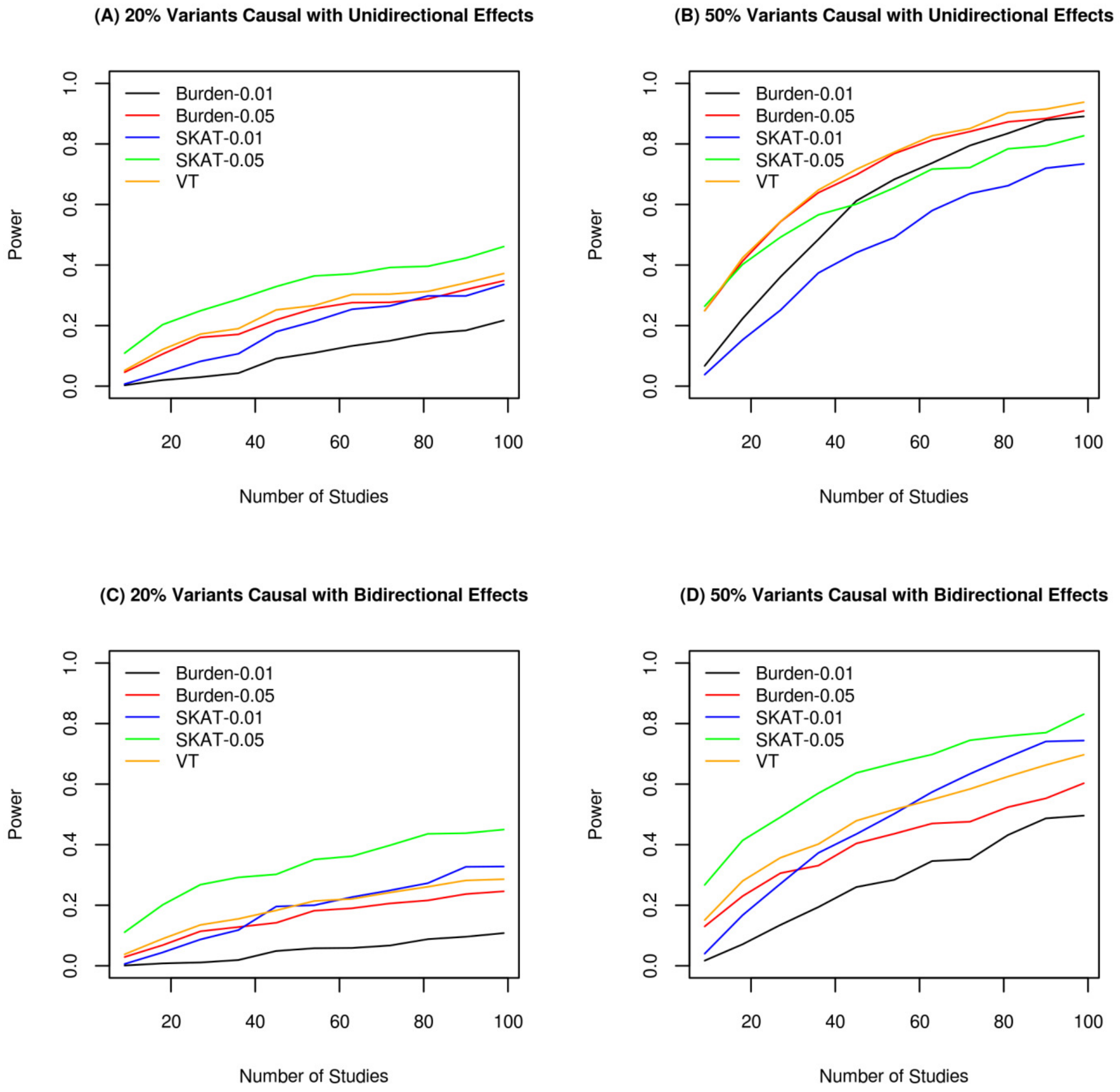

**Supplemental Figure 5**: Power of meta-analysis rare variant association tests as a function of the number of studies in the meta-analysis, each with 1000 individuals. Quantitative trait data were simulated

assuming the trait was influenced by rare variants with frequency less than 0.5%. In panel A, 20% of the variants with MAF <0.5% are causal and each increases trait values by 0.25 standard deviation units. In panel B, 50% of the variants with MAF <0.5% are causal and each increases trait values by 0.25 standard deviation units. In panel C, 20% of the variants with MAF < 0.5% are causal, with 80% of these increasing the trait by 0.25 standard deviation units with the other 20% decreasing trait values by the same amount. In panel D, 50% of the variants with MAF < 0.5% are causal, with 80% of these increasing the trait by 0.25 standard deviation units while the other 20% decreasing trait values by the same amount.

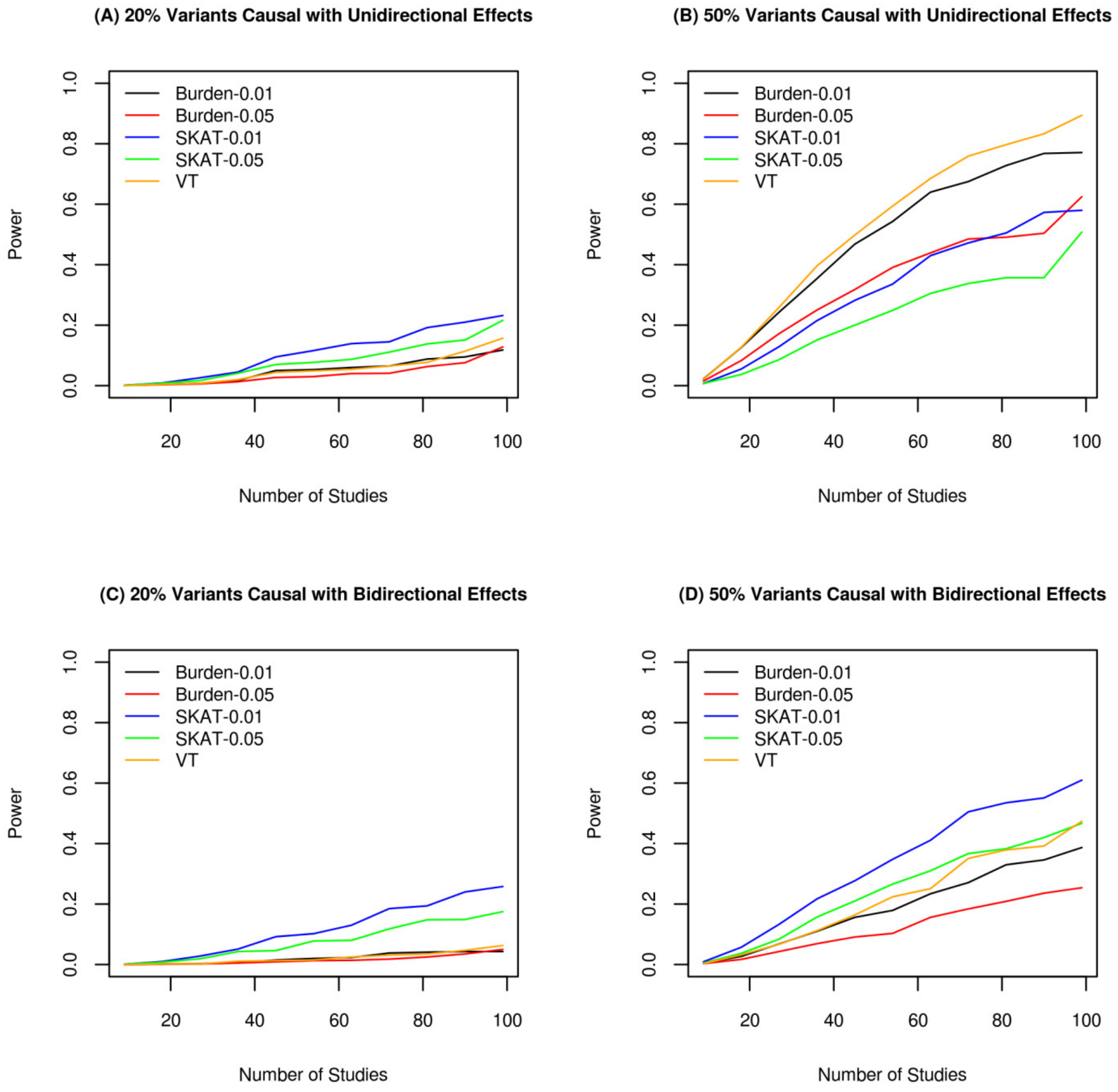

**Supplemental Figure 6:** Power comparison of meta-analysis and analysis of pooled individual level data ("Mega-Analysis"). Data were simulated for 20 samples of 1000 individuals each. Results are shown when (a) 20% of simulated variants are causal and each increase expected trait values by 0.25 standard deviation units; (b) 50% of the variants are causal and increase trait values by 0.25 standard deviation units; (c) 20% of variants are causal, with 80% increasing trait values by 0.25 standard deviation units and the remaining 20% decreasing trait values by the same amount; (d) 50% of variants are causal, with 80% increasing trait values by 0.25 standard deviation units and the other 20% decreasing trait values by the same amount.

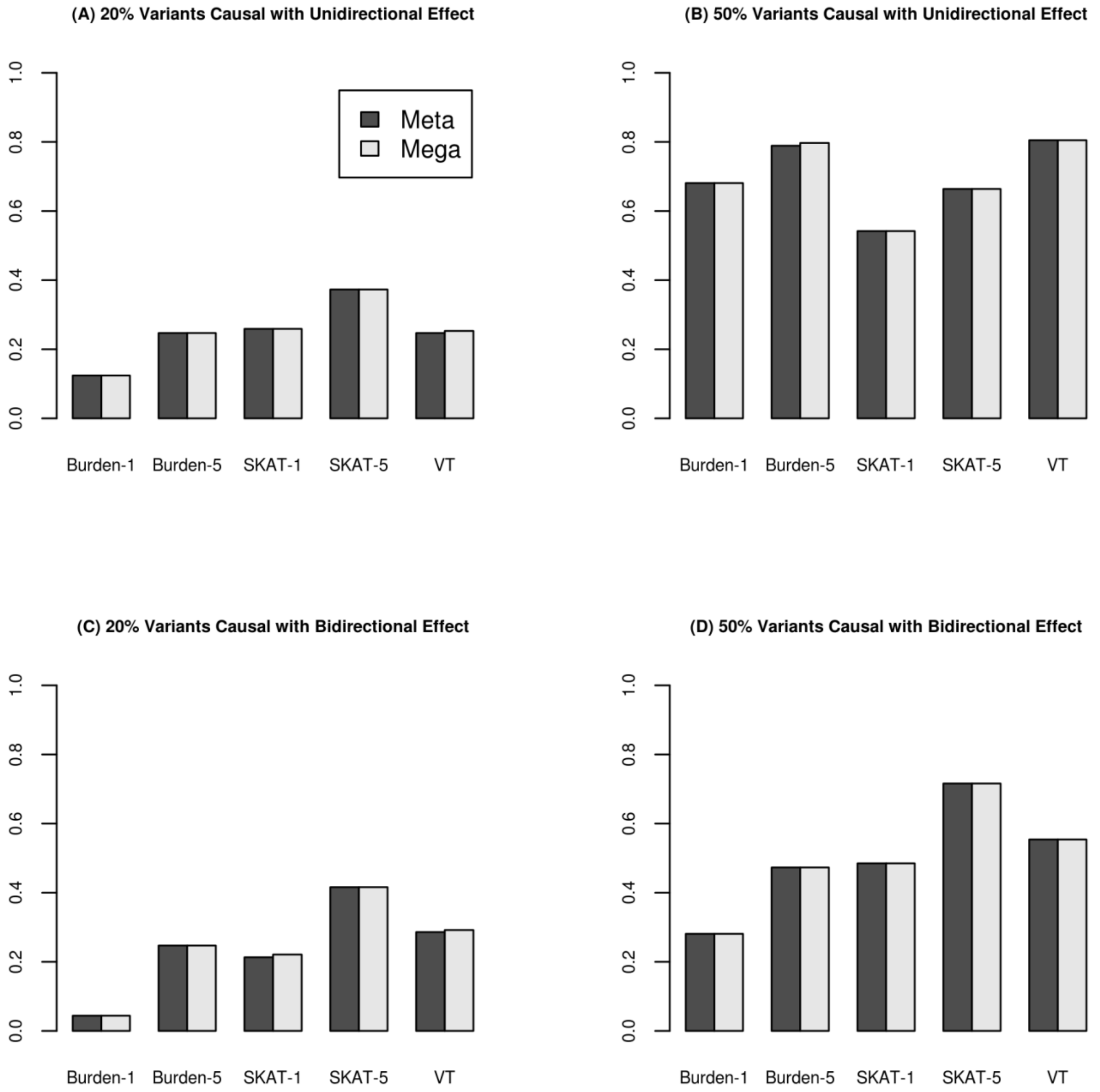

**Supplemental Figure 7:** Power comparison for our approach, Fisher's method and the minimal p-value approach. Three phenotype models were simulated: (A) half of low frequency variants with MAF < 0.5% are causal, each increasing expected trait values by 1/4 standard deviation; (B) half of all variants are causal, irrespective of frequency, and increase trait values by 1/4 standard deviation; (C) 50% of the variants are casual, irrespective of frequency, and 80% of these increase expected trait values by 1/4 standard deviation, while the remaining 20% decrease trait values by the same amount. Three samples of size 10,000 were simulated for each model, and meta-analysis was performed using either our approach or using Fisher's method and the minimal p-value approach to combine burden test, SKAT and variable threshold (VT) test statistics for variants with MAF<5%. The power was evaluated at the significance threshold of $\alpha=2.5\times10^{-6}$ using 10,000 replicates. Note that differences between our approach and these alternatives become more marked when sample sizes for individual studies differ and/or when more studies are meta-analyzed.

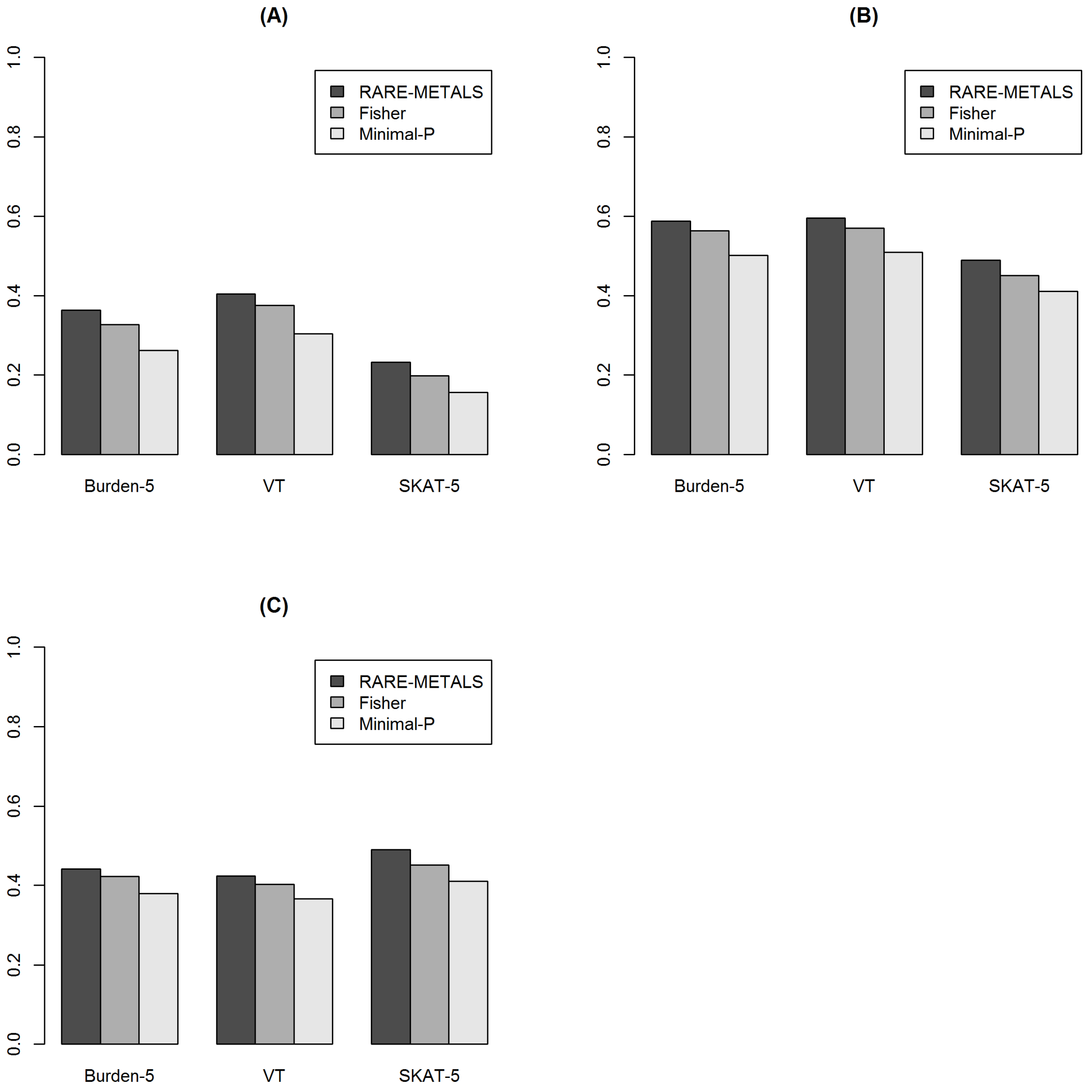

**Supplemental Figure 8**: Comparison of meta-analysis and analysis of pooled individual level data in the Malmö Diet and Cancer Study – Cardiovascular Cohort (MDC). Our largest sample, the MDC study was split into 4 sub-samples of comparable size. LDL-cholesterol values in each sub-sample were standardized using an inverse-normal transformation. Summary level statistics were generated for each sub-sample and meta-analyses was performed combining summary statistics from 4 studies. Results of the original analysis of the MDC study are also shown.

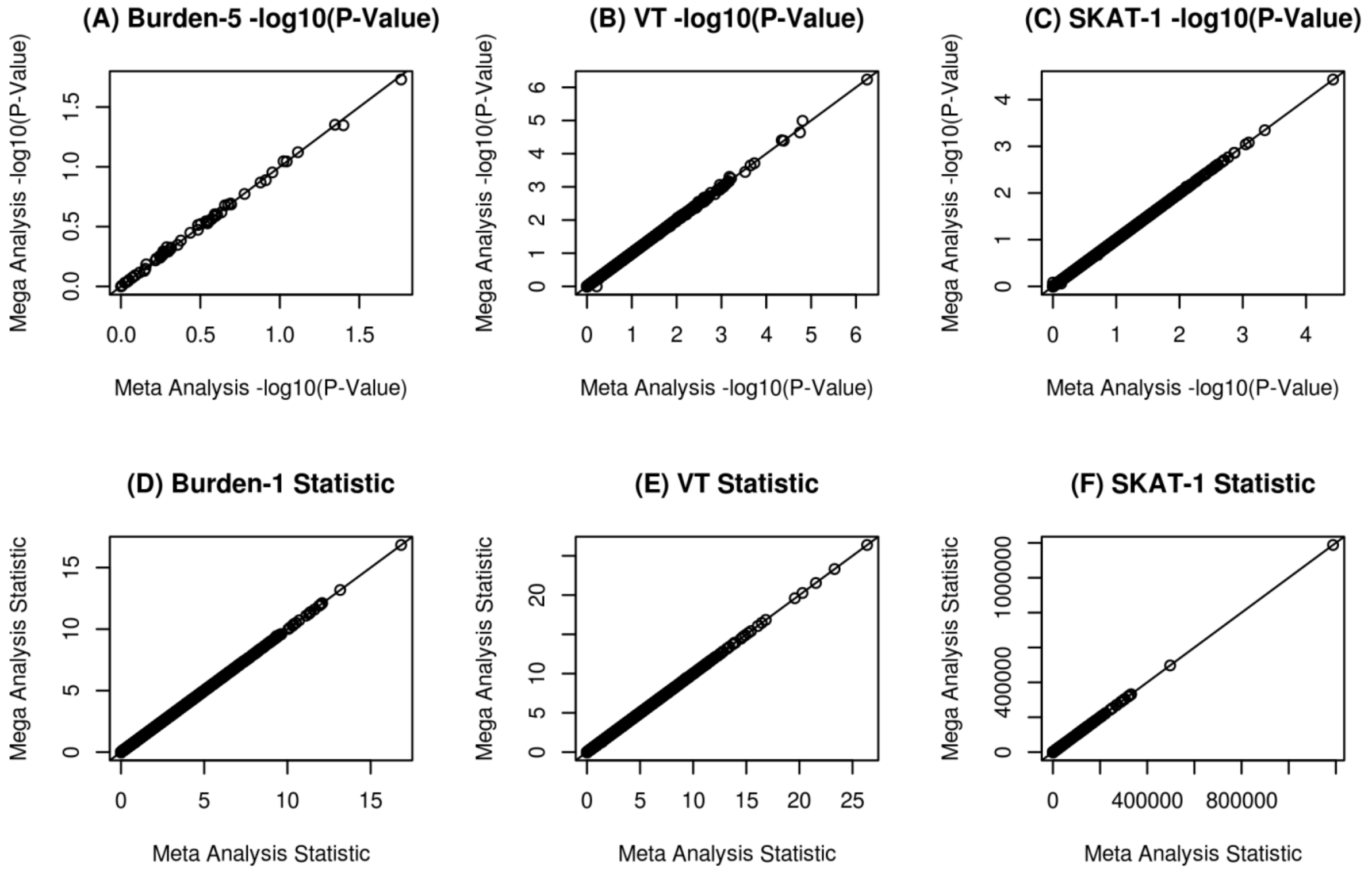

**Supplemental Figure 9**: Quantile-Quantile plot of p-values for single variant meta-analysis. Log-transformed observed and expected p-values are displayed for high density lipoprotein cholesterol (panels A-B), low density lipoprotein cholesterol (C-D) and triglyceride levels (E-F), either using all variants (left column) and variants with MAF < 5% (right column).

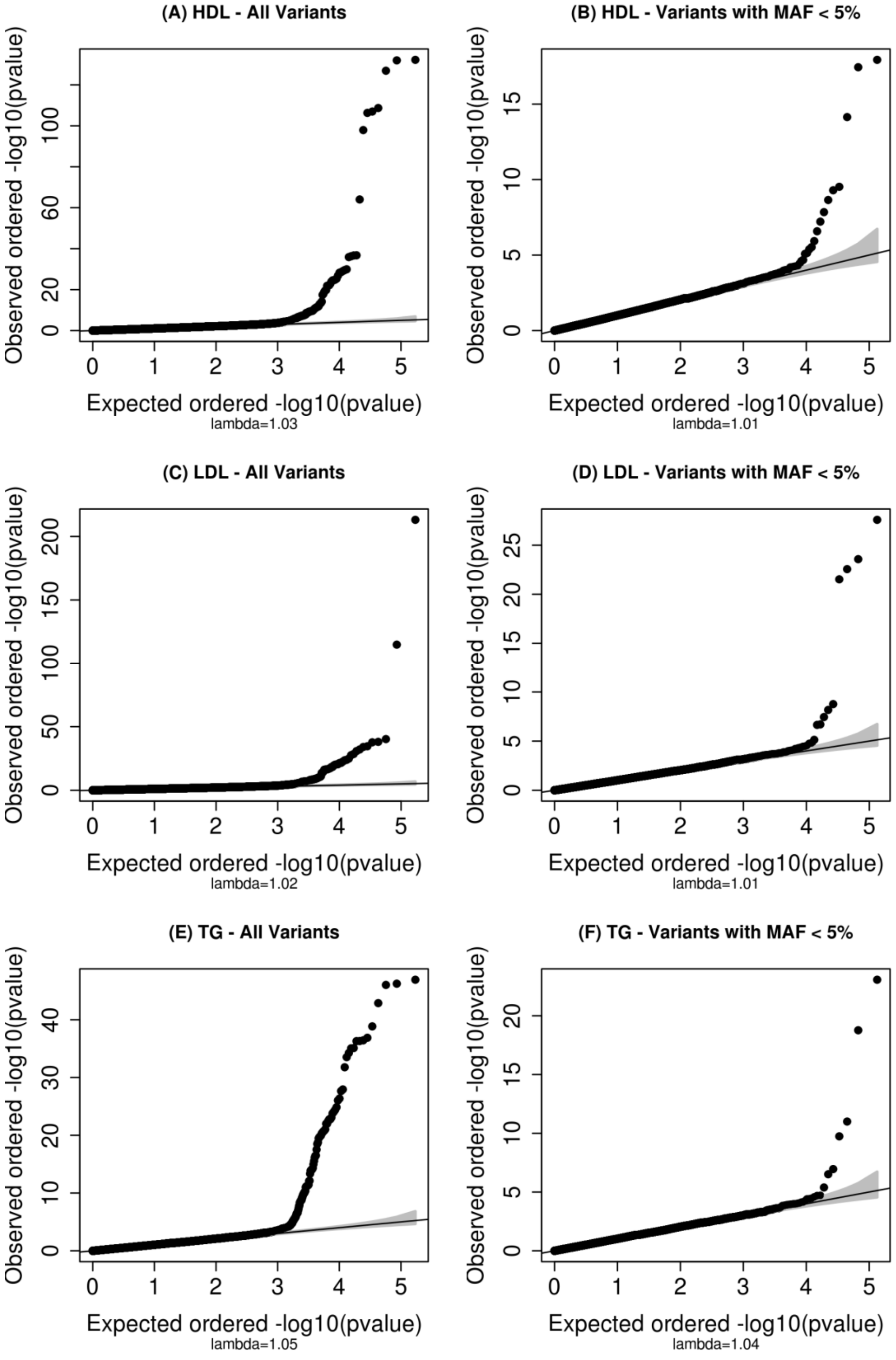

**Supplemental Figure 10 (Part 1 of 3):** Quantile-quantile plot of p-values for gene-level meta-analysis. Log-transformed observed and expected p-values are displayed for high density lipoprotein cholesterol (panels A-E), low density lipoprotein cholesterol (F-J) and triglyceride levels (K-O).

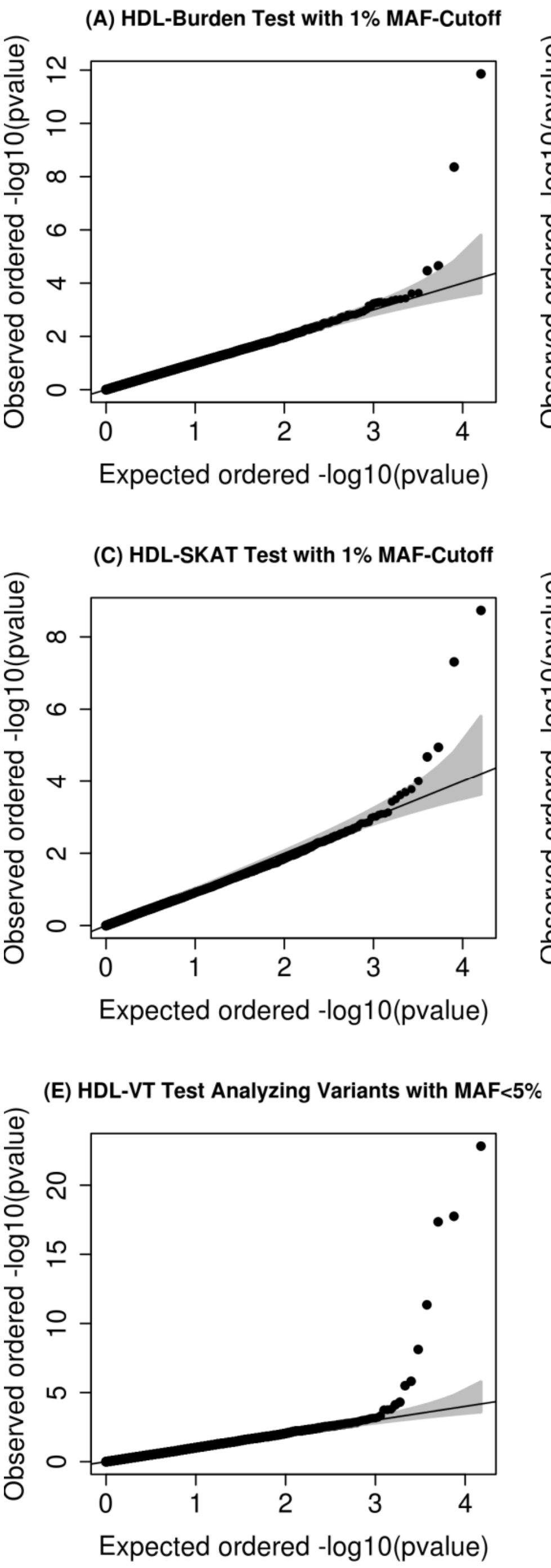

**Supplemental Figure 10 (Part 2 of 3):**

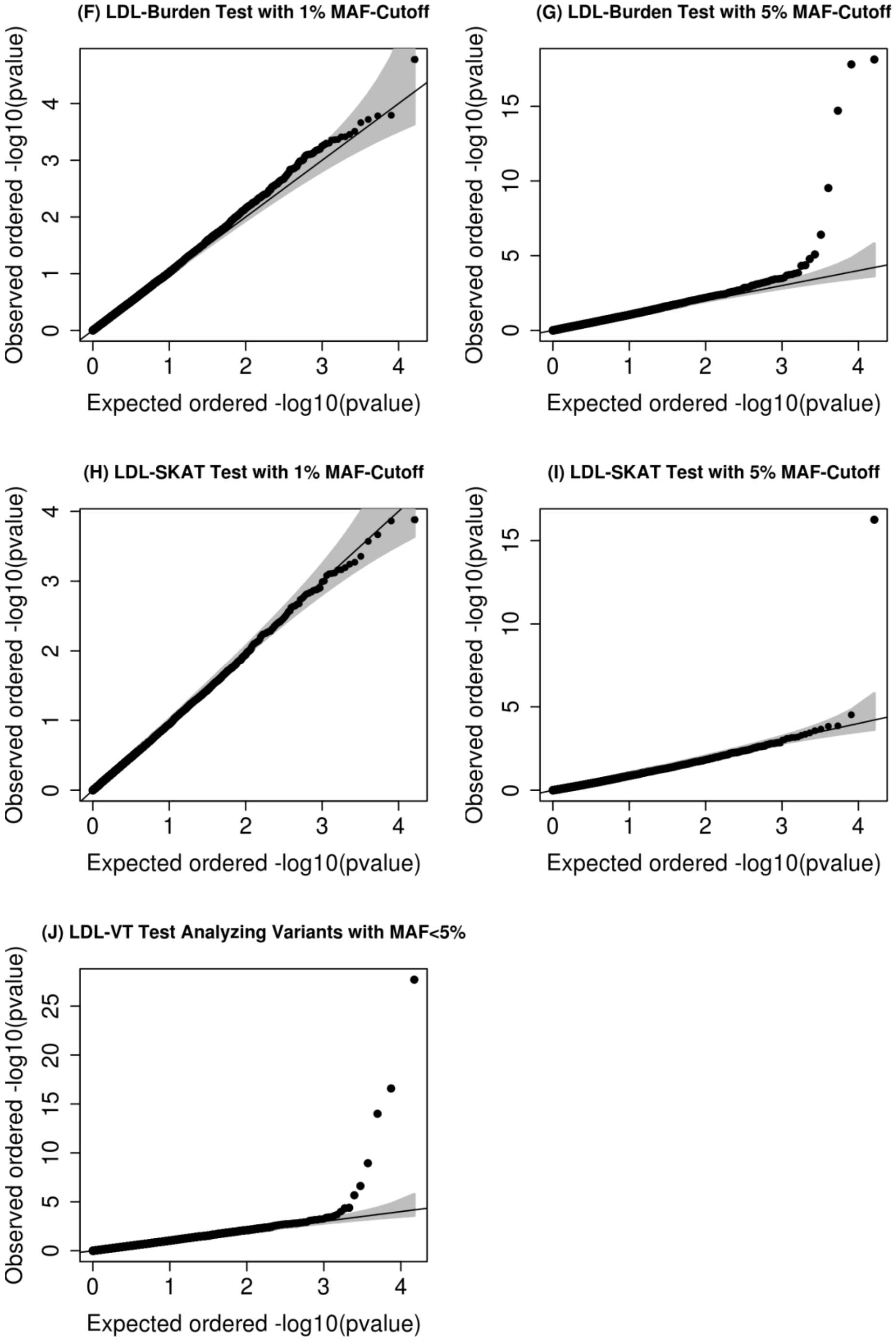

**Supplemental Figure 10 (Part 3 of 3):**

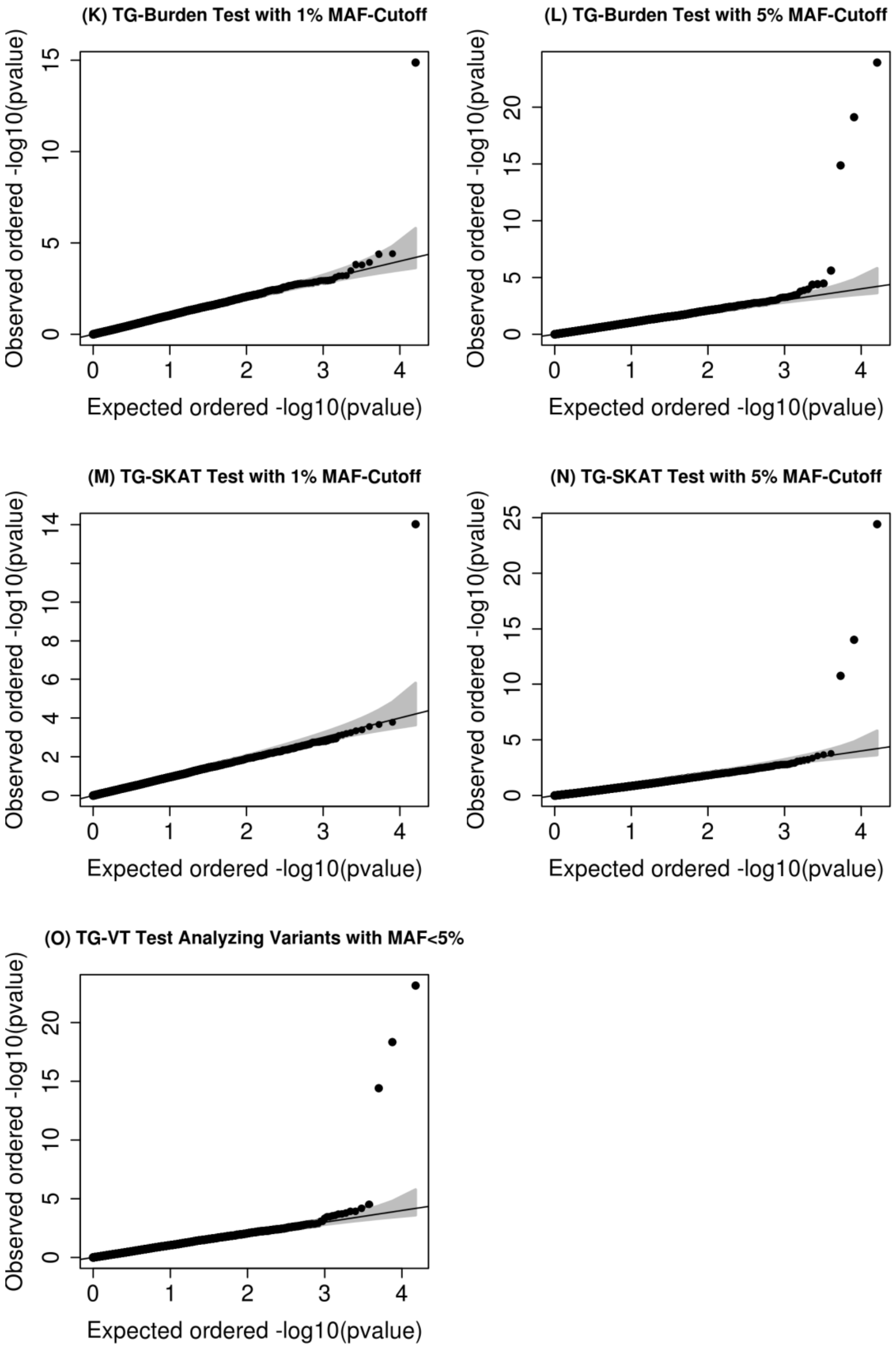

**Supplemental Figure 11**: Quantile-quantile plot for log-transformed p-values comparing the distribution of p-values in a conditional analysis (on the left) and an unconditional analysis (on the right). One hundred samples were simulated. In each simulation, a single common variant (with MAF>10%) was marked as causal, increasing expected trait values by 0.25 standard deviation units. A series of rare variant association analysis were then carried out, with (left) or without (right) conditioning on the effect of this common variant. The result clearly shows that, without conditioning, rare variant association test statistics are inflated.

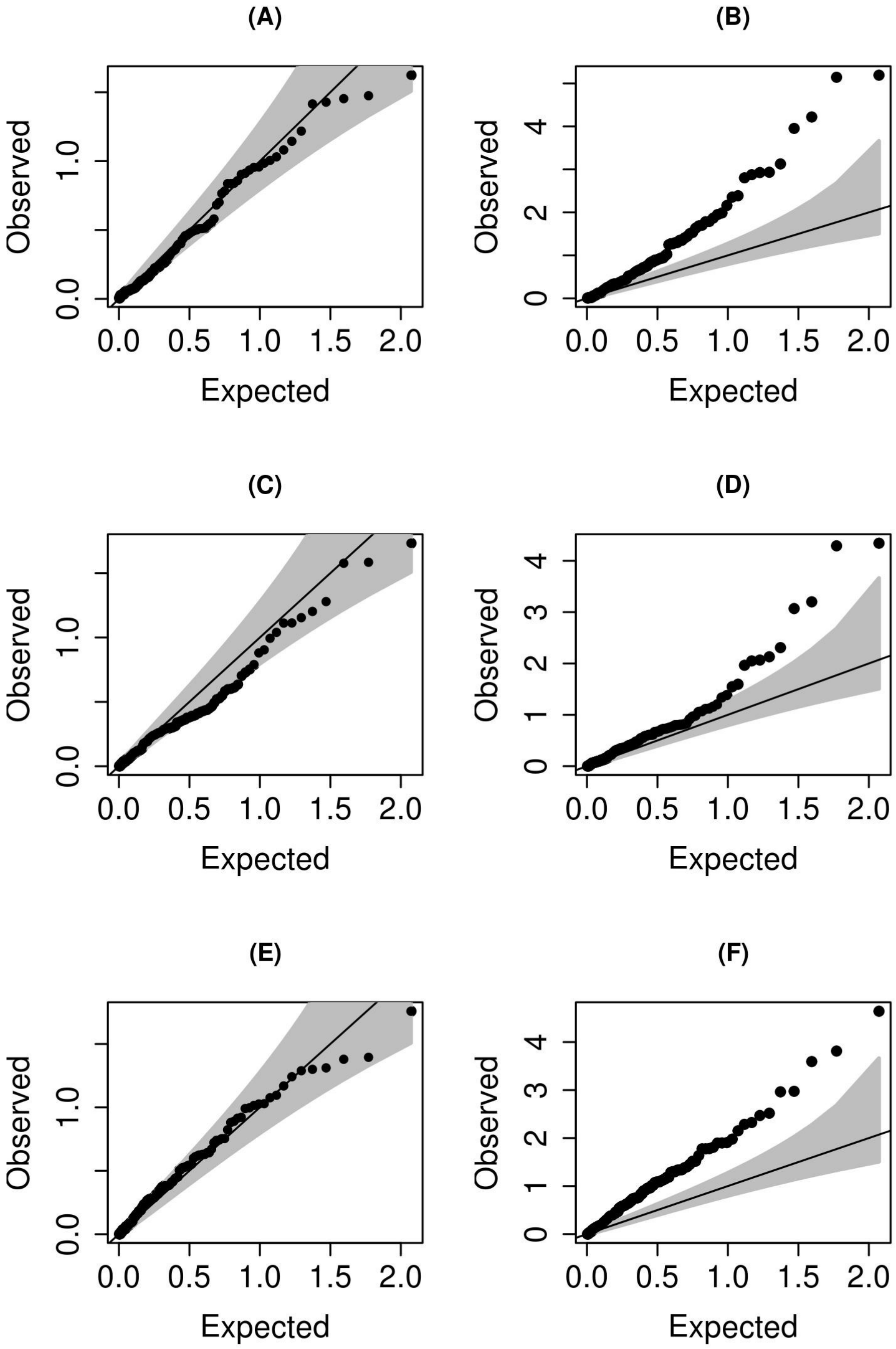

**Supplemental Figure 12**: Comparison of conditional meta-analysis and conditional analysis of pooled individual level data in the Malmö Diet and Cancer Study – Cardiovascular Cohort (MDC). Our largest sample, the MDC study was split into 4 sub-samples of comparable size. LDL-cholesterol values in each sub-sample were standardized using an inverse-normal transformation. Summary statistics were generated for each sub-sample. Conditional association analysis was performed in the pooled sample by controlling for the most significant single nucleotide polymorphism (rs7412) in gene *APOE*. Conditional evidence for association at sixty-six genes within 1Mb of rs7412 was evaluated, either using our meta-analysis approach (X axis) or by analyzing individual level data directly (Y axis). The genes examined were *TOMM40, APOE, OPA3, ERCC1, MARK4, FOSB, PVRL2, CKM, CLPTM1, RTN2, ZNF155, ZNF230, ZFP112, ZNF225, ZNF223, ZNF221, ZNF222, DMPK, ZNF45, CEACAM19, CLASRP, BCL3, EML2, SIX5, GEMIN7, PPP1R13L, FBXO46, PVR, CBLC, LOC100379224, ZNF227, ZNF235, ZNF285, CEACAM20, ZNF296, NKPD1, TRAPPC6A, BLOC1S3, KLC3, PPM1N, IRF2BP1, MYPOP, ERCC2, ZNF226, CD3EAP, GIPR, ZNF180, DMWD, BCAM, SYMPK, ZNF229, RSPH6A, ZNF234, VASP, APOC4, APOC1, FOXA3, EXOC3L2, RELB, ZNF224, APOC4-APOC2, APOC2, CEACAM16, ZNF284, QPCTL, ZNF233*.

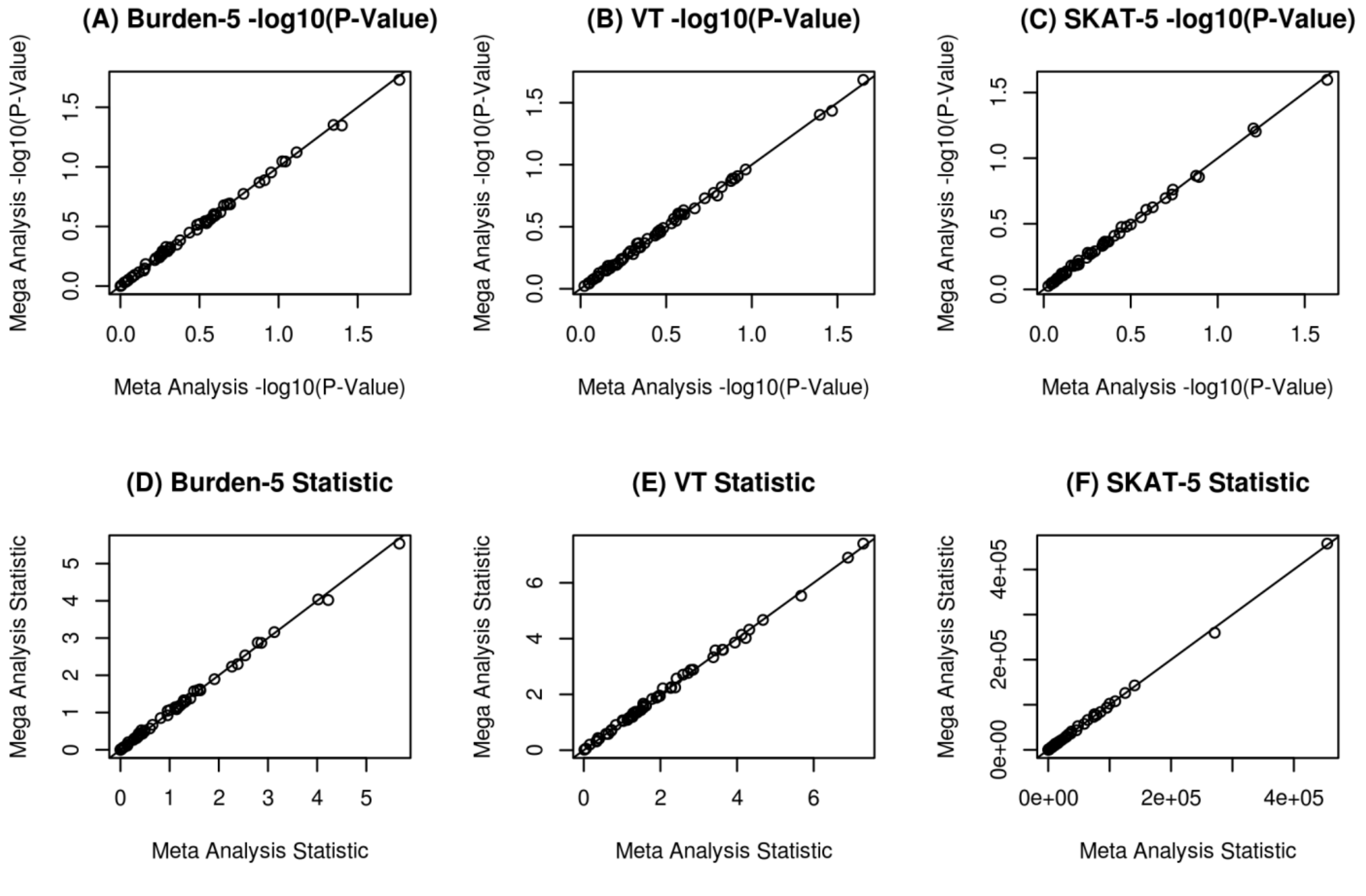

**Supplemental Figure 13:** Comparison of conditional meta-analysis and conditional analysis of pooled individual level data in the Malmö Diet and Cancer Study – Cardiovascular Cohort (MDC). Our largest sample, the MDC study was split into 4 sub-samples of comparable size. LDL-cholesterol values in each sub-sample were standardized using an inverse-normal transformation. Summary level statistics were generated for each sub-sample. C onditional association analysis was performed in the pooled sample conditional on three common single nucleotide polymorphism (rs6511720, rs2228671 and rs72658855) in gene *LDLR* that each reach significant evidence for association. A total of 59 genes within 1Mb of these top 3 SNPs were analyzed, either using our meta-analysis approach (X axis) or by analyzing individual level data directly (Y axis). Test statistics were evaluated at *DNM2, S1PR5, LOC55908, ZNF844, CCDC151, TYK2, ZNF440, ZNF491, CNN1, SLC44A2, SMARCA4, KEAP1, ICAM3, KANK2, ICAM1, RAVER1, EPOR, ICAM4, ZNF439, ILF3, DNMT1, PRKCSH, TMED1, KRI1, QTRT1, C19orf38, C19orf52, SPC24, TMEM205, C19orf39, ECSIT, ZNF441, ZNF69, ZNF700, ZNF433, AP1M2, TSPAN16, ACP5, RGL3, LDLR, YIPF2, CDKN2D, S1PR2, ZGLP1, ZNF653, LPPR2, DOCK6, PDE4A, ZNF627, CCDC159, ATG4D, RAB3D, CARM1, ICAM5, MRPL4, ZNF823, FDX1L, ZNF763, ZNF878.*

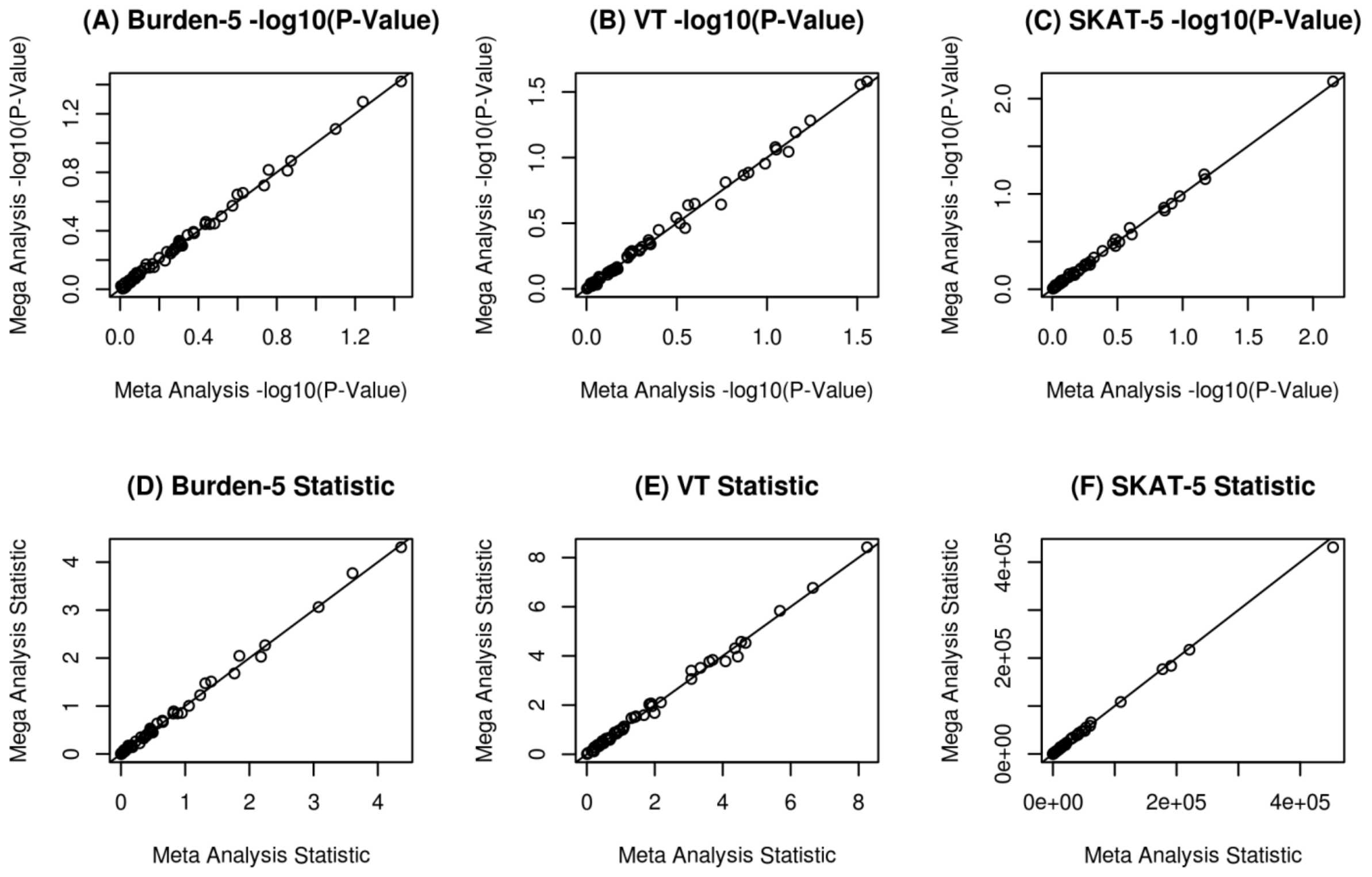

**Supplemental Table 1**: Evaluation of type I error rates for meta-analysis methods. Type I error rates were evaluated for three rare variant tests (burden-1: a simple burden test group variants with <1% frequency, VT: a variable threshold association test, SKAT-1: a sequence kernel association test focused on variants with frequency <1% and allowing for variants with opposite effects to reside in the same gene). Significance levels $\alpha$=0.001, 0.0001, and $2.5\times10^{-6}$ were considered. Data were generated for meta-analysis of 3, 6 and 9 samples of 1000 individuals. The type I error estimates are based upon $5\times10^{7}$ null simulations.

| Number of Studies | Burden-1 | VT | SKAT-1 |
|---|---|---|---|
| $\alpha=1\times10^{-3}$ | | | |
| 3 | $9.9\times10^{-4}$ | $1.0\times10^{-3}$ | $9.4\times10^{-4}$ |
| 6 | $1.0\times10^{-3}$ | $1.0\times10^{-3}$ | $9.8\times10^{-4}$ |
| 9 | $1.0\times10^{-3}$ | $1.0\times10^{-3}$ | $1.0\times10^{-3}$ |
| $\alpha=1\times10^{-4}$ | | | |
| 3 | $9.9\times10^{-5}$ | $1.1\times10^{-4}$ | $9.2\times10^{-5}$ |
| 6 | $1.1\times10^{-4}$ | $1.2\times10^{-4}$ | $9.9\times10^{-5}$ |
| 9 | $1.0\times10^{-4}$ | $1.1\times10^{-4}$ | $9.9\times10^{-5}$ |
| $\alpha=2.5\times10^{-6}$ | | | |
| 3 | $2.2\times10^{-6}$ | $2.6\times10^{-6}$ | $2.5\times10^{-6}$ |
| 6 | $2.4\times10^{-6}$ | $2.6\times10^{-6}$ | $2.6\times10^{-6}$ |
| 9 | $2.2\times10^{-6}$ | $1.6\times10^{-6}$ | $2.2\times10^{-6}$ |

**Supplemental Table 2:** Summary trait and variant information. In each study, medians and interquartile ranges are tabulated for age, sex and lipid traits, together with the number of genotyped non-synonymous and loss-of-function variants. Participating studies were the Malmo Diet and Cancer (MDC) study, the Ottawa Heart study, the Women's Health Initiative Sequencing Project (WHISP), Procardis and HUNT. Genotyped samples in Procardis and HUNT are separated into heart disease cases and controls.

| **Study** | **Age** | **HDL** | **LDL** | **TG** | **Total Number of Individuals** | **Proportion of Males** | **Number of Variants** | | |
|---|---|---|---|---|---|---|---|---|---|
| | **Median (Interquartile Range) Lipids Level (mg/dL)** | | | | | | All | Nonsynonymous + Loss of Function | Nonsynonymous + Loss of Function, MAF <1% |
| **Malmo Diet and Cancer** | 58 (10.4) | 51.4 (18.9) | 158.3 (50.2) | 102.7 (65.5) | 4924 | 40.8% | 130,621 | 111,127 | 90,317 |
| **Ottawa Heart Study** | 72 (13.1) | 50.6 (20.9) | 139.4 (48.3) | 121.2 (100.0) | 2938 | 60.0% | 116,173 | 97,628 | 77,866 |
| **WHISP European Americans** | 68 (7.0) | 56.0 (22.0) | 140.2 (48.1) | 139.0 (87.0) | 2031 | 0.0% | 110,678 | 91,998 | 70,421 |
| **PROCARDIS (Cases)** | 58 (10.0) | 46.0 (17.0) | 142.0 (54.3) | 159.0 (120.0) | 2070 | 48.9% | 97,887 | 79,551 | 58,864 |
| **PROCARDIS (Controls)** | 66 (7.0) | 53.9 (20.5) | 129.9 (42.2) | 123.1 (88.5) | 1299 | 62.1% | 105,255 | 86,639 | 66,114 |
| **HUNT (Cases)** | 66(18.0) | 46.3(15.4) | 162.2(57.9) | 177 (123.9) | 2659 | 65.5% | 90,340 | 72,866 | 52,133 |
| **HUNT (Controls)** | 65(19.0) | 50.2(19.3) | 154.4(54.1) | 141.6 (86.7) | 2778 | 66.1% | 91,902 | 74,366 | 53,682 |

**Supplemental Table 3** Variants sites shared between studies, by frequency. The number of shared variant nucleotide sites are displayed respectively for each pair of studies and for variant sites with MAF > 1% and with MAF≤ 1%. Tabulated studies include the Malmo Diet and Cancer (MDC) study, the Ottawa Heart Study (Ottawa), European American Samples from the WHISP study, and case and control samples from Procardis and HUNT.

| | MDC | Ottawa | WHISP European Americans | PROCARDIS (Cases) | PROCARDIS (Controls) | HUNT (Cases) | HUNT (Controls) |
|---|---|---|---|---|---|---|---|
| | | | **Variants with MAF > 1%** | | | | |
| **MDC** | 36,520 | 33,546 | 34,414 | 34,925 | 34,903 | 34,263 | 34,251 |
| **Ottawa** | | 34,981 | 34,472 | 34,311 | 34,265 | 33,178 | 33,183 |
| **WHISP European Americans** | | | 37,256 | 35,090 | 35,020 | 33,942 | 33,929 |
| **PROCARDIS (Cases)** | | | | 36,484 | 35,609 | 34,370 | 34,390 |
| **PROCARDIS (Controls)** | | | | | 36,283 | 34,317 | 34,327 |
| **HUNT (Cases)** | | | | | | 36,140 | 35,516 |
| **HUNT (Controls)** | | | | | | | 36,098 |
| | | | **Variants with MAF ≤ 1%** | | | | |
| **MDC** | 94,101 | 57,932 | 53,613 | 48,682 | 53,051 | 44,583 | 45,773 |
| **Ottawa** | | 81,192 | 55,396 | 47,528 | 52,283 | 39,414 | 40,493 |
| **WHISP European Americans** | | | 73,422 | 43,892 | 47,858 | 36,627 | 37,614 |
| **PROCARDIS (Cases)** | | | | 61,403 | 45,837 | 35,867 | 36,670 |
| **PROCARDIS (Controls)** | | | | | 68,972 | 38,204 | 39,061 |
| **HUNT (Cases)** | | | | | | 54,200 | 45,054 |
| **HUNT (Controls)** | | | | | | | 55,804 |

**Supplemental Table 4**: Results for single variant meta-analysis. Loci that are statistically significant after Bonferroni correction (with $p < 3\times10^{-7}$) are shown. In each locus, p-value, annotation, reference and alternative allele, alternative allele frequency as well as genetic effect estimate and standard deviation are displayed for the variant with the most significant p-value.

| **Gene** | **Gene Position[a]** | **P-value** | **rs#** | **Annotation** | **Ref/Alt** | **Frequency for Alt Allele** | **Estimated Effect Size for Alt Allele (in standard deviation units)** | **Standard Error for Estimated Effect (in standard deviation units)** |
|---|---|---|---|---|---|---|---|---|
| | | | | **HDL** | | | | |
| ***LPL*** | chr8:19.8Mb | $1.17\times10^{-18}$ | rs268 | Nonsynonymous | A/G | 0.02454 | -0.2963 | 0.001129 |
| ***ANGPTL4*** | chr19:8.4Mb | $3.61\times10^{-18}$ | rs116843064 | Nonsynonymous | G/A | 0.02649 | 0.2809 | 0.001045 |
| ***LIPG*** | chr18:47.1Mb | $7.26\times10^{-15}$ | rs77960347 | Nonsynonymous | A/G | 0.01346 | 0.3484 | 0.002006 |
| ***CD300LG*** | chr17:41.9Mb | $3.00\times10^{-10}$ | rs72836561 | Nonsynonymous | C/T | 0.03327 | -0.1815 | 0.00083 |
| ***LIPC*** | chr15:58.9Mb | $5.10\times10^{-10}$ | rs113298164 | Nonsynonymous | C/T | 0.003675 | 0.5356 | 0.007425 |
| ***APOB*** | chr2:21.2Mb | $2.24\times10^{-9}$ | rs533617 | Nonsynonymous | T/C | 0.0397 | 0.1592 | 0.000709 |
| ***HNF4A*** | chr20:43.0Mb | $2.64\times10^{-7}$ | rs1800961 | Nonsynonymous | C/T | 0.041 | -0.1342 | 0.00068 |
| | | | | **LDL** | | | | |
| ***PCSK9*** | chr1:55.5Mb | $2.60\times10^{-28}$ | rs11591147 | Nonsynonymous | G/T | 0.01311 | -0.5112 | 0.002146 |
| ***BCAM*** | chr19:45.3Mb | $2.64\times10^{-24}$ | rs28399653 | Nonsynonymous | G/A | 0.03509 | -0.2896 | 0.00081 |
| ***CBLC*** | chr19:45.3Mb | $2.99\times10^{-22}$ | rs3208856 | Nonsynonymous | C/T | 0.03678 | -0.27 | 0.000774 |
| ***PVR*** | chr19:45.2Mb | $1.71\times10^{-9}$ | rs1058402 | Nonsynonymous | G/A | 0.0487 | -0.1458 | 0.000586 |
| ***APOB*** | chr2:21.2Mb | $6.70\times10^{-9}$ | rs5742904 | Nonsynonymous | C/T | 0.000633 | 1.211 | 0.043597 |
| | | | | **TG** | | | | |
| ***ANGPTL4*** | chr19:8.4Mb | $8.55\times10^{-24}$ | rs116843064 | Nonsynonymous | G/A | 0.0265 | -0.3248 | 0.001043 |
| ***LPL*** | chr8:19.8Mb | $1.70\times10^{-19}$ | rs268 | Nonsynonymous | A/G | 0.02472 | 0.3022 | 0.00112 |
| ***APOB*** | chr2:21.2Mb | $1.81\times10^{-10}$ | rs533617 | Nonsynonymous | T/C | 0.03967 | -0.1697 | 0.000708 |
| ***MAP1A*** | chr15:43.8Mb | $1.10\times10^{-7}$ | rs55707100 | Nonsynonymous | C/T | 0.026 | 0.1732 | 0.001065 |

a. Gene position is defined based upon hg19, GRCh37 Genome Reference Consortium Human Reference 37

**Supplemental Table 5 (Part 1 of 3):** Comparison of meta-analysis and analysis of individual studies for gene-level tests. Results for six rare variant tests are shown (burden-5: a simple burden test group variants with <5% or <1% frequency, VT: a variable threshold association test, SKAT-5: a sequence kernel association test focused on variants with <5% or <1% frequency and allowing for variants with opposite effects to reside in the same gene). For tests that assume that model the average effect of variants in a gene, a + or – sign indicates whether these variants raised (+) or lowered (-) trait levels on average. Overall, the results show that, for these genes, meta-analysis results in a substantially stronger signal than analysis of any single sample and that the direction of effect for these top signals is generally consistent across studies. Study abbreviations are as in previous tables.

| Gene | Meta Analysis | MDC | Ottawa | WHISP | PROCARD IS (Cases) | PROCARD IS (Controls) | HUNT (Cases) | HUNT (Controls) |
|---|---|---|---|---|---|---|---|---|
| | | | | **Burden-5** | | | | |
| | | | | **HDL** | | | | |
| ***LPL*** | $2\times10^{-24}$/- | $5\times10^{-11}$/- | $4\times10^{-5}$/- | 0.007/- | $5\times10^{-4}$/- | 0.004/- | 0.002/- | 0.001/- |
| ***ANGPTL 4*** | $3\times10^{-19}$/+ | $2\times10^{-5}$/+ | $2\times10^{-6}$/+ | 0.04/+ | 0.1/+ | 0.006/+ | 0.03/+ | $4\times10^{-6}$/+ |
| ***LIPG*** | $6\times10^{-19}$/+ | $1\times10^{-8}$/+ | 0.03/+ | 0.2/+ | 0.003/+ | 0.04/+ | $5\times10^{-7}$/+ | 0.001/+ |
| ***HNF4A*** | $3\times10^{-7}$/- | 0.009/- | $5\times10^{-4}$/- | 0.003/- | 0.7/+ | 0.002/- | 0.8/- | 0.08/- |
| ***LIPC*** | $4\times10^{-7}$/+ | $8\times10^{-4}$/+ | 0.4/+ | 0.4/+ | 0.5/+ | 0.3/+ | 0.007/+ | $9\times10^{-4}$/+ |
| ***CD300LG*** | $8\times10^{-7}$/- | 0.04/- | 0.002/- | 0.1/- | 0.8/+ | 0.09/- | 0.7/- | $1\times10^{-4}$/- |
| | | | | **LDL** | | | | |
| ***PCSK9*** | $7\times10^{-19}$/- | $5\times10^{-5}$/- | $1\times10^{-9}$/- | $5\times10^{-7}$/- | 0.02/- | 0.001/- | 0.02/- | 0.06/- |
| ***BCAM*** | $2\times10^{-18}$/- | $6\times10^{-6}$/- | 0.4/- | 0.02/- | 0.01/- | 0.01/- | 0.03/- | $3\times10^{-6}$/- |
| ***CBLC*** | $2\times10^{-15}$/- | $3\times10^{-7}$/- | 0.1/- | 0.3/- | 0.2/- | $5\times10^{-5}$/- | 0.002/- | $2\times10^{-4}$/- |
| ***PVR*** | $3\times10^{-10}$/- | $2\times10^{-5}$/- | 0.02/- | 0.05/- | 0.01/- | 0.2/- | 0.1/+ | 0.3/- |
| | | | | **TG** | | | | |
| ***ANGPTL 4*** | $1\times10^{-24}$/- | $3\times10^{-6}$/- | $9\times10^{-6}$/- | $2\times10^{-6}$/- | 0.01/- | 0.005/- | 0.002/- | $2\times10^{-6}$/- |
| ***LPL*** | $8\times10^{-20}$/+ | $1\times10^{-9}$/+ | 0.001/+ | $5\times10^{-4}$/+ | 0.001/+ | 0.2/+ | 0.04/+ | $5\times10^{-5}$/+ |

**Supplemental Table 5 (Part 2 of 3).**

| Gene | Meta Analysis | MDC | Ottawa | WHISP | PROCARD IS (Cases) | PROCARD IS (Controls) | HUNT (Cases) | HUNT (Controls) |
|---|---|---|---|---|---|---|---|---|
| | | | | **SKAT-5** | | | | |
| | | | | **HDL** | | | | |
| ***ANGPTL 4*** | $3\times10^{-19}$/+ | $5\times10^{-5}$/+ | $2\times10^{-6}$/+ | 0.03/+ | 0.02/+ | 0.03/+ | 0.04/+ | $7\times10^{-6}$/+ |
| ***LPL*** | $5\times10^{-13}$/- | $8\times10^{-6}$/- | 0.003/- | 0.08/- | 0.07/- | 0.1/- | 0.1/- | 0.05/- |
| ***LIPG*** | $3\times10^{-9}$/+ | $3\times10^{-4}$/+ | 0.04/+ | 0.4/+ | 0.03/+ | 0.2/+ | 0.002/+ | 0.07/+ |
| ***HNF4A*** | $3\times10^{-7}$/- | 0.009/- | $4\times10^{-4}$/- | 0.003/- | 0.8/+ | 0.003/- | 0.8/- | 0.08/- |
| | | | | **LDL** | | | | |
| ***PCSK9*** | $6\times10^{-17}$/- | $8\times10^{-4}$/- | $4\times10^{-7}$/- | $3\times10^{-9}$/- | 0.003/- | $3\times10^{-5}$/- | 0.2/- | 0.1/- |
| | | | | **TG** | | | | |
| ***ANGPTL 4*** | $4\times10^{-25}$/- | $4\times10^{-6}$/- | $8\times10^{-6}$/- | $8\times10^{-6}$/- | 0.006/- | 0.02/- | 0.005/- | $1\times10^{-5}$/- |
| ***LPL*** | $2\times10^{-11}$/+ | $5\times10^{-7}$/+ | 0.02/+ | 0.02/+ | 0.1/+ | 0.3/+ | 0.2/+ | 0.02/+ |

**Supplemental Table 5 (Part 3 of 3).**

| | | | | | | | | |
|---|---|---|---|---|---|---|---|---|
| | | | | **VT** | | | | |
| | | | | **HDL** | | | | |
| ***LPL*** | $1\times10^{-23}$/- | $1\times10^{-10}$/- | $2\times10^{-4}$/- | 0.02/- | 0.002/- | 0.02/- | 0.006/- | 0.003/- |
| ***ANGPTL 4*** | $2\times10^{-18}$/+ | $1\times10^{-4}$/+ | $2\times10^{-5}$/+ | 0.1/+ | 0.4/- | 0.02/+ | 0.07/+ | $2\times10^{-5}$/+ |
| ***LIPG*** | $4\times10^{-18}$/+ | $6\times10^{-8}$/+ | 0.04/- | 0.4/- | 0.01/+ | 0.1/+ | $1\times10^{-6}$/+ | 0.003/+ |
| ***LIPC*** | $4\times10^{-12}$/+ | 0.003/- | 0.2/- | 0.5/- | 0.8/- | 0.7/- | $1\times10^{-4}$/- | $4\times10^{-7}$/- |
| | | | | **LDL** | | | | |
| ***PCSK9*** | $2\times10^{-28}$/- | $2\times10^{-5}$/- | $3\times10^{-9}$/- | $2\times10^{-6}$/- | 0.001/- | $9\times10^{-6}$/- | 0.08/- | 0.008/- |
| ***BCAM*** | $3\times10^{-17}$/- | $9\times10^{-5}$/- | 0.7/- | 0.007/- | 0.01/- | 0.003/- | 0.07/- | $4\times10^{-6}$/- |
| ***CBLC*** | $1\times10^{-14}$/- | $7\times10^{-7}$/- | 0.4/- | 0.06/+ | 0.06/- | $3\times10^{-4}$/- | 0.01/- | 0.001/- |
| ***PVR*** | $1\times10^{-9}$/- | $8\times10^{-5}$/- | 0.08/- | 0.2/- | 0.06/- | 0.3/- | 0.3/+ | 0.7/+ |
| ***LDLR*** | $2\times10^{-7}$/+ | 0.1/- | 0.001/- | 0.2/- | $5\times10^{-4}$/- | 0.1/- | 0.03/- | 0.008/- |
| | | | | **TG** | | | | |
| ***ANGPTL 4*** | $7\times10^{-24}$/- | $1\times10^{-5}$/- | $5\times10^{-5}$/- | $6\times10^{-6}$/- | 0.04/+ | 0.01/- | 0.005/- | $7\times10^{-6}$/- |
| ***LPL*** | $5\times10^{-19}$/+ | $4\times10^{-9}$/+ | 0.006/+ | 0.001/+ | 0.005/+ | 0.6/+ | 0.06/- | $2\times10^{-4}$/+ |
| | | | | **Burden-1** | | | | |
| | | | | **HDL** | | | | |
| ***LIPC*** | $1\times10^{-12}$/+ | 0.004/+ | 0.05/+ | 0.2/+ | 0.7/+ | 0.8/+ | $4\times10^{-5}$/+ | $1\times10^{-7}$/+ |
| | | | | **SKAT-1** | | | | |
| | | | | **HDL** | | | | |
| ***LIPC*** | $2\times10^{-9}$/+ | 0.07/+ | 0.1/+ | 0.08/+ | 0.7/+ | 0.8/+ | 0.001/+ | $3\times10^{-5}$/+ |

**Supplemental Table 6:** Comparison of gene-level test results with single variant association tests. For each locus identified using gene-level association tests, we show the rs number, ref/alt allele, alt allele frequency and p-value for the variant site that displays the most significant p-value. The loci where one or more gene-based association signal exceeds the top single variant association signal are labeled with an asterisk.

| Gene | Burden-1 | Burden-5 | SKAT-1 | SKAT-5 | VT | MAF Cutoff | Top Single Variant Association(MAF<5%) | | | |
|---|---|---|---|---|---|---|---|---|---|---|
| | | | | | | | rs Number | Ref/Alt | p-value | AF |
| | | | | | **HDL** | | | | | |
| ***LIPC**** | **$1.4\times10^{-12}$** | **$3.5\times10^{-7}$** | **$1.8\times10^{-9}$** | $1.4\times10^{-2}$ | **$4.5\times10^{-12}$** | $3.7\times10^{-3}$ | rs113298164 | C/T | $5.1\times10^{-10}$ | $3.68\times10^{-3}$ |
| ***LPL**** | 0.97 | **$2.5\times10^{-24}$** | 0.35 | **$5.0\times10^{-13}$** | **$1.5\times10^{-23}$** | 0.025 | rs268 | A/G | $1.2\times10^{-18}$ | 0.025 |
| ***ANGPTL4**** | 0.022 | **$2.9\times10^{-19}$** | 0.022 | **$3.0\times10^{-19}$** | **$1.8\times10^{-18}$** | 0.026 | rs116843064 | G/A | $3.6\times10^{-18}$ | 0.027 |
| ***LIPG**** | $2.2\times10^{-5}$ | **$6.4\times10^{-19}$** | $2.1\times10^{-5}$ | **$2.9\times10^{-9}$** | **$4.4\times10^{-18}$** | 0.013 | rs77960347 | A/G | $7.3\times10^{-15}$ | 0.014 |
| ***HNF4A*** | 0.74 | **$2.8\times10^{-7}$** | 0.68 | **$2.5\times10^{-7}$** | **$1.5\times10^{-6}$** | 0.041 | rs1800961 | C/T | $2.6\times10^{-7}$ | 0.041 |
| ***CD300LG*** | 0.49 | **$8.5\times10^{-7}$** | 0.52 | $1.0\times10^{-5}$ | **$3.1\times10^{-6}$** | 0.033 | rs72836561 | C/T | $3.0\times10^{-10}$ | 0.033 |
| | | | | | **LDL** | | | | | |
| ***PCSK9**** | $1.8\times10^{-2}$ | **$7.4\times10^{-19}$** | 0.081 | **$5.5\times10^{-17}$** | **$2.0\times10^{-28}$** | 0.013 | rs11591147 | G/T | $2.6\times10^{-28}$ | 0.013 |
| ***BCAM*** | 0.17 | **$1.6\times10^{-18}$** | 0.15 | $3.0\times10^{-5}$ | **$2.6\times10^{-17}$** | 0.036 | rs28399653 | G/A | $2.6\times10^{-24}$ | 0.035 |
| ***CBLC*** | 0.94 | **$2.0\times10^{-15}$** | 0.44 | $1.5\times10^{-4}$ | **$1.0\times10^{-14}$** | 0.044 | rs3208856 | C/T | $3.0\times10^{-22}$ | 0.037 |
| ***PVR*** | 0.061 | **$3.0\times10^{-10}$** | 0.048 | 0.063 | **$1.1\times10^{-9}$** | 0.049 | rs1058402 | G/A | $1.7\times10^{-9}$ | 0.049 |
| ***LDLR**** | $1.8\times10^{-3}$ | $4.7\times10^{-5}$ | 0.038 | 0.25 | **$2.4\times10^{-7}$** | $5.2\times10^{-4}$ | rs139791325 | G/A | $7.68\times10^{-4}$ | $5.2\times10^{-4}$ |
| | | | | | **TG** | | | | | |
| ***ANGPTL4**** | 0.026 | **$1.2\times10^{-24}$** | 0.037 | **$3.9\times10^{-25}$** | **$7.1\times10^{-24}$** | 0.026 | rs116843064 | G/A | $8.6\times10^{-24}$ | 0.027 |
| ***LPL**** | 0.68 | **$7.7\times10^{-20}$** | 0.26 | **$1.8\times10^{-11}$** | **$4.6\times10^{-19}$** | 0.025 | rs268 | A/G | $1.7\times10^{-19}$ | 0.025 |

**Supplemental Table 7**: Results of conditional association analysis for trait LDL and variants in gene *LDLR*. We performed conditional association analysis for variants in *LDLR*, conditioning on 3 common variants (rs6511720, rs2228671 and rs72658855) that are statistically significant in single variant association analysis (i.e. with p-value $<3\times10^{-7}$). The rs number, reference, alternative alleles, minor allele frequencies, p-values before and after conditioning on top variants, the estimates of genetic effects for alternative alleles, and annotation information are displayed for non-synonymous and loss-of-function variants.

| Single Variant Association Analysis | | | | | | | | |
|---|---|---|---|---|---|---|---|---|
| **RS** | **Ref** | **Alt** | **MAF** | **Un-conditional P-value** | **Conditional P-value** | **Un-conditional Beta[a]** | **Conditional Beta[a]** | **Anno** |
| rs6511720 | G | T | 0.1083 | $2\times10^{-38}$ | - | -0.22 | - | Intron |
| rs2228671 | C | T | 0.1145 | $4\times10^{-22}$ | - | -0.16 | - | Synonymous |
| rs2738459 | A | C | 0.4908 | $4\times10^{-8}$ | - | -0.06 | - | Intron |
| rs11669576 | G | A | 0.046 | $8\times10^{-4}$ | 0.201 | 0.08 | 0.03 | Nonsynonymous |
| rs139624145 | G | A | 0.0001 | $8\times10^{-4}$ | 0.001 | 1.68 | 1.61 | Nonsynonymous |
| rs139791325 | G | A | 0.0005 | $8\times10^{-4}$ | 0.002 | 0.77 | 0.7 | Nonsynonymous |
| rs199774121 | C | A | $3\times10^{-5}$ | 0.004 | 0.002 | 2.88 | 3.14 | Stop_Gain |
| rs144172724 | G | A | $3\times10^{-5}$ | 0.024 | 0.037 | 2.26 | 2.11 | Nonsynonymous |
| rs141673997 | G | A | $3\times10^{-5}$ | 0.048 | 0.056 | 1.98 | 1.92 | Nonsynonymous |
| rs150673992 | C | T | $6\times10^{-5}$ | 0.056 | 0.031 | 1.35 | 1.54 | Nonsynonymous |
| rs28942084 | C | T | $6\times10^{-5}$ | 0.151 | 0.158 | 1.02 | 1.01 | Nonsynonymous |
| rs139043155 | T | A | 0.0001 | 0.21 | 0.241 | 0.63 | 0.59 | Nonsynonymous |
| rs139361635 | G | A | $3\times10^{-5}$ | 0.266 | 0.2 | 1.11 | 1.29 | Nonsynonymous |
| rs143992984 | G | A | $8\times10^{-5}$ | 0.358 | 0.233 | 0.53 | 0.7 | Nonsynonymous |
| rs137853963 | G | A | 0.0018 | 0.391 | 0.212 | -0.11 | -0.16 | Nonsynonymous |
| rs13306505 | C | T | $6\times10^{-5}$ | 0.511 | 0.47 | 0.47 | 0.51 | Nonsynonymous |
| rs148698650 | G | A | 0.0001 | 0.539 | 0.585 | 0.27 | 0.25 | Nonsynonymous |
| rs200727689 | G | A | $3\times10^{-5}$ | 0.603 | 0.667 | 0.52 | 0.43 | Nonsynonymous |
| rs5928 | G | A | $3\times10^{-5}$ | 0.892 | 0.851 | -0.14 | -0.19 | Nonsynonymous |
| rs146200173 | C | G | 0.0002 | 0.997 | 0.99 | 0 | 0 | Nonsynonymous |

| Gene-level Test | | | | | | |
|---|---|---|---|---|---|---|
| **Gene** | **P-value before Conditioning** | **P-value after Conditioning** | **MAF cutoff before conditioning** | **MAF cutoff after conditioning** | **Estimate of Genetic Effect Before Conditioning[a]** | **Estimate of Genetic Effect After Conditioning[a]** |
| ***LDLR*** | $2.4\times10^{-7}$ | $4.6\times10^{-7}$ | $5.2\times10^{-4}$ | $5.2\times10^{-4}$ | 0.75 | 0.73 |

a. Estimates are calculated in standard deviation units.

**Supplemental Table 8**: Results of conditional association analysis for trait LDL and locus *APOE-BCAM-CBLC-PVR*. We performed conditional association analysis conditioning on the top variant (rs7412) that are statistically significant in single variant association analysis (i.e. with p-value $<3\times10^{-7}$). The p-values before and after conditional analysis for burden test and SKAT tests with 5% MAF cutoff and VT test that analyzes variant with MAF<5% are shown. The rs number, reference, alternative alleles, p-values before and after conditioning on rs7412 were also displayed for each gene.

| Gene | Burden-5 | | SKAT-5 | | VT | | Top SNP | | | | |
|---|---|---|---|---|---|---|---|---|---|---|---|
| | Un-conditional | Conditional | Un-conditional | Conditiona l | Un-conditional | Conditiona l | RS | Ref | Alt | Un-conditional | Conditional |
| ***BCAM*** | $1.57\times10^{-18}$ | 0.89 | $3.01\times10^{-5}$ | 0.42 | $2.61\times10^{-17}$ | 0.80 | rs28399653 | G | A | $2.64\times10^{-24}$ | 0.67 |
| ***CBLC*** | $1.98\times10^{-15}$ | 0.02 | $1.47\times10^{-4}$ | 0.41 | $9.99\times10^{-15}$ | 0.09 | rs3208856 | C | T | $2.99\times10^{-22}$ | 0.76 |
| ***PVR*** | $2.97\times10^{-10}$ | 0.14 | $6.30\times10^{-2}$ | 0.62 | $1.13\times10^{-9}$ | 0.39 | rs1058402 | G | A | $1.71\times10^{-9}$ | 0.27 |